\documentclass[prl, aps,
 twocolumn,
 superscriptaddress,
 amsmath,amssymb,
 10pt
]{revtex4-2}

\usepackage{times}
\usepackage{color}
\usepackage{graphicx}
\usepackage[dvipsnames]{xcolor}
\usepackage{physics}
\usepackage{bm}
\usepackage{mathtools}
\usepackage{upgreek}
\usepackage[colorlinks=true,allcolors=blue]{hyperref}
\usepackage{footmisc}
\usepackage{makecell}
\usepackage{mathtools}
\usepackage{soul}
\usepackage{xcolor}
\usepackage{lipsum}
\usepackage{wasysym}
\usepackage{array} 
\usepackage{titlesec}
\titleformat{\section} 
{\normalfont\Large\bfseries}{\thesection}{1em}{} 
\titleformat{\subsection} 
	{\normalfont\bfseries}{\makebox[30pt][l]{\thesubsection}}{0pt}{}
\titleformat{\subsubsection} 
	{\normalfont \itshape}{\makebox[30pt][l]{\thesubsection}}{0pt}{}

\soulregister{\cite}{7}
\soulregister{\ref}{7}
\soulregister{\eqref}{7}
\soulregister{\label}{7}
\soulregister{\onlinecite}{7}
\definecolor{mycolor}{HTML}{ffffd4}
\sethlcolor{mycolor}

\begin{document}

\title{Identifying Materials-Level Sources of Performance Variation in Superconducting Transmon Qubits}

\author{Akshay A. Murthy$^*$}
\affiliation{Superconducting Quantum Materials and Systems Division, Fermi National Accelerator Laboratory (FNAL), Batavia, IL 60510, USA}
\email{Corresponding author: amurthy@fnal.gov}

\author{Mustafa Bal}
\affiliation{Superconducting Quantum Materials and Systems Division, Fermi National Accelerator Laboratory (FNAL), Batavia, IL 60510, USA}
\author{Michael J. Bedzyk}
\affiliation{Department of Materials Science and Engineering, Northwestern University, Evanston, IL, 60208, USA}
\author{Hilal Cansizoglu}
\affiliation{Rigetti Computing, Berkeley, CA 94710, USA}%
\author{Randall K. Chan}
\affiliation{Ames National Laboratory, Ames, IA 50011, USA}
\affiliation{Department of Physics \& Astronomy, Iowa State University, Ames, IA 50011, USA}
\author{Venkat Chandrasekhar}
\affiliation{Department of Physics and Astronomy, Northwestern University, Evanston, IL 60208, USA}
\author{Francesco Crisa}
\affiliation{Superconducting Quantum Materials and Systems Division, Fermi National Accelerator Laboratory (FNAL), Batavia, IL 60510, USA}
\author{Amlan Datta}
\affiliation{Ames National Laboratory, Ames, IA 50011, USA}
\affiliation{Department of Physics \& Astronomy, Iowa State University, Ames, IA 50011, USA}
\author{Yanpei Deng}
\affiliation{Department of Physics and Astronomy, Northwestern University, Evanston, IL 60208, USA}
\author{Celeo D. Matute Diaz}
\affiliation{Department of Materials Science and Engineering, Northwestern University, Evanston, IL, 60208, USA}
\author{Vinayak P. Dravid}
\affiliation{Department of Materials Science and Engineering, Northwestern University, Evanston, IL, 60208, USA}
\affiliation{The NU\textit{ANCE} Center, Northwestern University, Evanston, IL, 60208, USA}
\affiliation{International Institute of Nanotechnology, Northwestern University, Evanston, IL, 60208, USA}
\author{David A. Garcia-Wetten}
\affiliation{Department of Materials Science and Engineering, Northwestern University, Evanston, IL, 60208, USA}
\author{Sabrina Garattoni}
\affiliation{Superconducting Quantum Materials and Systems Division, Fermi National Accelerator Laboratory (FNAL), Batavia, IL 60510, USA}
\author{Sunil Ghimire}
\affiliation{Ames National Laboratory, Ames, IA 50011, USA}
\affiliation{Department of Physics \& Astronomy, Iowa State University, Ames, IA 50011, USA}
\author{Dominic P. Goronzy}
\affiliation{Department of Materials Science and Engineering, Northwestern University, Evanston, IL, 60208, USA}
\author{Sebastian E. de Graaf}
\affiliation{National Physical Laboratory, Teddington TW11 0LW, United Kingdom}
\author{Sam Haeuser}
\affiliation{Ames National Laboratory, Ames, IA 50011, USA}
\affiliation{Department of Physics \& Astronomy, Iowa State University, Ames, IA 50011, USA}
\author{Mark C. Hersam}
\affiliation{Department of Materials Science and Engineering, Northwestern University, Evanston, IL, 60208, USA}
\affiliation{Department of Chemistry, Northwestern University, Evanston, IL 60208, USA}
\affiliation{Department of Electrical and Computer Engineering, Northwestern University, Evanston, IL 60208, USA}
\author{Peter Hopkins}
\affiliation{National Institute of Standards and Technology, Boulder, CO, USA}
\author{Dieter Isheim}
\affiliation{Department of Materials Science and Engineering, Northwestern University, Evanston, IL, 60208, USA}
\author{Kamal Joshi}
\affiliation{Ames National Laboratory, Ames, IA 50011, USA}
\author{Richard Kim}
\affiliation{Ames National Laboratory, Ames, IA 50011, USA}
\author{Saagar Kolachina}
\affiliation{Ames National Laboratory, Ames, IA 50011, USA}
\author{Cameron J. Kopas}
\affiliation{Rigetti Computing, Berkeley, CA 94710, USA}
\author{Matthew J. Kramer}
\affiliation{Ames National Laboratory, Ames, IA 50011, USA}
\author{Ella O. Lachman}
\affiliation{Rigetti Computing, Berkeley, CA 94710, USA}
\author{Jaeyel Lee}
\affiliation{Superconducting Quantum Materials and Systems Division, Fermi National Accelerator Laboratory (FNAL), Batavia, IL 60510, USA}
\author{Peter G. Lim}
\affiliation{Applied Physics Graduate Program, Northwestern University, Evanston, IL, 60208, USA}
\author{Andrei Lunin}
\affiliation{Superconducting Quantum Materials and Systems Division, Fermi National Accelerator Laboratory (FNAL), Batavia, IL 60510, USA}
\author{William Mah}
\affiliation{Department of Materials Science and Engineering, Northwestern University, Evanston, IL, 60208, USA}
\author{Jayss Marshall}
\affiliation{Rigetti Computing, Berkeley, CA 94710, USA}
\author{Josh Y. Mutus}
\affiliation{Rigetti Computing, Berkeley, CA 94710, USA}
\author{Jin-Su Oh}
\affiliation{Ames National Laboratory, Ames, IA 50011, USA}
\author{David Olaya}
\affiliation{National Institute of Standards and Technology, Boulder, CO, USA}
\affiliation{Department of Physics, University of Colorado, Boulder, CO, USA}
\author{David P. Pappas}
\affiliation{Rigetti Computing, Berkeley, CA 94710, USA}
\author{Joong-mok Park}
\affiliation{Ames National Laboratory, Ames, IA 50011, USA}
\author{Ruslan Prozorov}
\affiliation{Ames National Laboratory, Ames, IA 50011, USA}
\affiliation{Department of Physics \& Astronomy, Iowa State University, Ames, IA 50011, USA}
\author{Roberto dos Reis}
\affiliation{Department of Materials Science and Engineering, Northwestern University, Evanston, IL, 60208, USA}
\affiliation{The NU\textit{ANCE} Center, Northwestern University, Evanston, IL, 60208, USA}
\author{David N. Seidman}
\affiliation{Department of Materials Science and Engineering, Northwestern University, Evanston, IL, 60208, USA}
\author{Zuhawn Sung}
\affiliation{Superconducting Quantum Materials and Systems Division, Fermi National Accelerator Laboratory (FNAL), Batavia, IL 60510, USA}
\author{Makariy Tanatar}
\affiliation{Ames National Laboratory, Ames, IA 50011, USA}
\author{Mitchell J. Walker}
\affiliation{Department of Materials Science and Engineering, Northwestern University, Evanston, IL, 60208, USA}
\author{Jigang Wang}
\affiliation{Ames National Laboratory, Ames, IA 50011, USA}
\author{Maxwell Wisne}
\affiliation{Department of Physics and Astronomy, Northwestern University, Evanston, IL 60208, USA}
\author{Haotian Wu}
\affiliation{Department of Materials Science and Engineering, Iowa State University, Ames, IA 50011, USA}
\author{Lin Zhou}
\affiliation{Ames National Laboratory, Ames, IA 50011, USA}
\affiliation{Department of Materials Science and Engineering, Iowa State University, Ames, IA 50011, USA}
\author{Shaojiang Zhu}
\affiliation{Superconducting Quantum Materials and Systems Division, Fermi National Accelerator Laboratory (FNAL), Batavia, IL 60510, USA}
\author{Anna Grassellino}
\affiliation{Superconducting Quantum Materials and Systems Division, Fermi National Accelerator Laboratory (FNAL), Batavia, IL 60510, USA}
\author{Alexander Romanenko}
\affiliation{Superconducting Quantum Materials and Systems Division, Fermi National Accelerator Laboratory (FNAL), Batavia, IL 60510, USA}

\begin{abstract}
The Superconducting Materials and Systems (SQMS) Center, a DOE National Quantum Information Science Research Center, has conducted a comprehensive and coordinated study using superconducting transmon qubit chips with known performance metrics to identify the underlying materials-level sources of device-to-device performance variation. Following qubit coherence measurements, these qubits of varying base superconducting metals and substrates have been examined with various nondestructive and invasive material characterization techniques at Northwestern University, Ames National Laboratory, and Fermilab as part of a blind study. We find trends in variations of the depth of the etched substrate trench, the thickness of the surface oxide, and the geometry of the sidewall, which when combined, lead to correlations with the T$_1$ lifetime across different devices. In addition, we provide a list of features that varied from device to device, for which the impact on performance requires further studies. Finally, we identify two low-temperature characterization techniques that may potentially serve as proxy tools for qubit measurements. These insights provide materials-oriented solutions to not only reduce performance variations across neighboring devices, but also to engineer and fabricate devices with optimal geometries to achieve performance metrics beyond the state-of-the-art values.

\end{abstract}

\maketitle
Advances in our understanding of defects, impurities, interfaces, and surfaces associated with superconducting materials have played a crucial role in driving recent increases in coherence times and gate fidelities in transmon qubits \cite{RN1, RN38, RN39, RN40, PhysRevApplied.19.024013, RN42, tuokkola2024methodsachievenearmillisecondenergy}. This includes both (1) identifying disordered regions that produce two-level system defects (TLS) and/or introduce pair breaking mechanisms that generate quasiparticles \cite{RN23,simmonds2004decoherence, Muller_2019, mcdermott2009materials, pritchard2024}, as well as (2) developing new strategies to mitigate their impact. Examples include encapsulating the surfaces of superconducting films to eliminate lossy oxides \cite{RN1, RN43, chang2024eliminatingsurfaceoxidessuperconducting} and developing annealing processes to reduce the two-level systems present in Josephson junctions \cite{RN4, RN44}. However, continued scaling of superconducting quantum platforms requires further advances in coherence time, not only at a single-qubit level but also reproducibly across a multiqubit chip \cite{mohseni2024buildquantumsupercomputerscaling, MURRAY2021100646, tuokkola2024methodsachievenearmillisecondenergy}.

To this end, with groups worldwide reporting wide variations between devices in the median T$_1$ for their best fabrication processes, reducing performance spread in qubit coherence remains a challenge in scaling to multi-qubit platforms. As an example, \citeauthor{RN1} reported that the median T$_1$ times of neighboring qubits with a common geometry ranged from 215 $\mu$s to 323 $\mu$s at frequencies that ranged between 3.8 GHz and 3.9 GHz within their highest performing chip \cite{RN1}. Similarly, in the case of the best performing chip measured by \citeauthor{RN40}, the group observed that median T$_1$ times of neighboring qubits with a common geometry ranged from 312 $\mu$s to 476 $\mu$s at frequencies between 3.8 GHz and 3.9 GHz \cite{RN40}. \citeauthor{RN38} observed similar spreads as they observed median T$_1$ times of qubits with common geometries and common processes ranging from 152 to 270 $\mu$s at frequencies between 2.8 GHz and 3.2 GHz \cite{RN38}. Finally, although \citeauthor{RN37} were successful in manufacturing superconducting transmon qubits using industrial fabrication methods as part of a 300mm complementary metal - oxide - semiconductor (CMOS) pilot line, the median T$_1$ times for the best performing qubit in each die ranged from roughly 42 $\mu$s at the edge of the 300mm wafer to 113 $\mu$s near the center \cite{RN37}. Given the complex geometries and various surfaces and interfaces present, which are made possible through a series of specialized deposition, lithography, and etching processes, this variation from qubit to qubit is not necessarily surprising. However, without greater control over this device-to-device variation, the achievable circuit depth and overall performance of multiqubit processors are greatly limited \cite{krantz2019quantum, mohseni2024buildquantumsupercomputerscaling}.

To gain insight into the primary factors giving rise to performance variations in neighboring devices, we performed a coordinated, blind study deploying a wide variety of materials and superconducting characterization techniques, as well as electromagnetic simulations to examine sources of loss in state-of-the-art transmon qubits prepared on both silicon and sapphire substrates. Our findings indicate trends in variations of the depth of the etched substrate trench, the thickness of the surface oxide, and the geometry of the sidewall, which when combined, lead to correlations with overall microwave loss across different devices. These observations allow us to build a hierarchy of losses, summarizing the relative impact of various defective structures that serve as sources of dissipation, while also providing a pathway to reduce performance variations across neighboring devices.

\section{Structure of Research Study}
\begin{figure}
    {\includegraphics[width =\columnwidth]{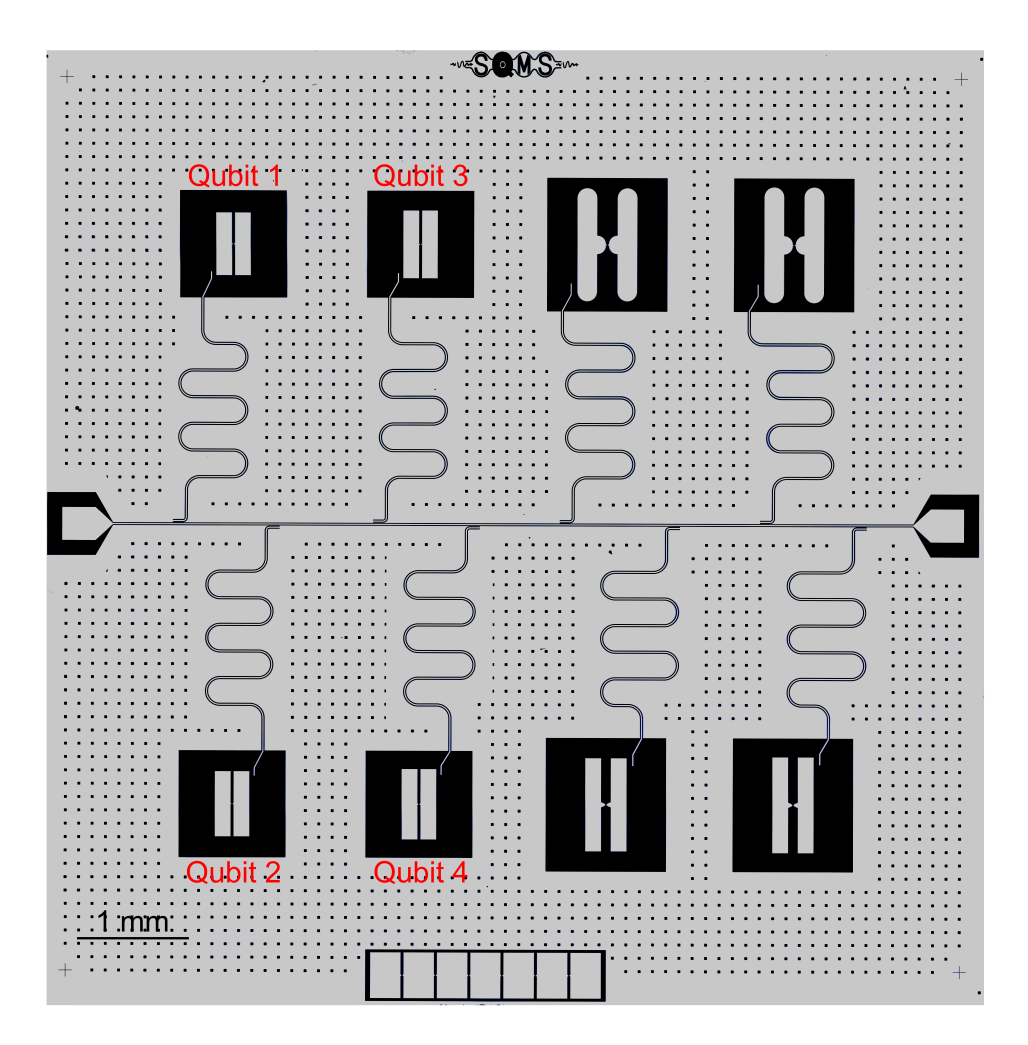}}
   \caption{8 qubit chip layout consisting of three different geometries. For the purpose of this study, qubits with a common geometry (Qubits 1-4) were analyzed with a variety of structural, chemical, and physical characterization techniques.}
   \label{fig:figure1}
\end{figure}

In this study, 22 superconducting transmon qubits across 7 chips were first measured and characterized according to the protocol described in the Supporting Information. T$_1$ measurements were performed continuously for 10 hrs for each qubit. The average T$_1$, the qubit frequency ($\omega_q$) and the average quality factor ($Q = 2\pi\omega_qT_1$) are provided in Table 1 of the qubit devices seen in Figure \ref{fig:figure1}. The spread in T$_1$ measurements for each qubit is provided in Figure \ref{fig:SQubit}. Given that both material quality and radiofrequency (RF) design impact the rate of qubit decoherence \cite{krantz2019quantum}, this study was carefully designed to specifically uncover materials-level sources of performance variation. By comparing and contrasting qubits of identical geometry and similar frequencies measured during a single cooldown one after another, we were able to keep extrinsic factors associated with the measurement setup, such as coupling strength to readout resonators and package and stray radiation, largely fixed. We also do not observe any positional dependence on the qubit T$_1$ as seen in Table 1.

All qubit devices employed a niobium base layer; a subset of devices additionally has a capping consisting of Ta, Au, or AuPd. As seen in Table 1, three sets of qubits were prepared on a sapphire substrate, while the remaining four were prepared on a silicon substrate. By exploring qubits consisting of a variety of materials, we aimed to glean generalizable relationships between materials and performance that are relevant to the community at large. Following qubit measurements, the devices were distributed for structure, chemical, and physical characterization at Ames National Laboratory, Northwestern University, and Fermilab. As part of the blind study, the qubit results were not revealed to the characterization team until after the characterization measurements were completed. The qubits were first probed with a series of non-destructive characterization methods, followed by a series of more invasive techniques. The materials-level causes of device variations were identified by probing the distinctive characteristics associated with each qubit device. The following are some of the structural, chemical, and physical properties and features that were measured and tracked for each of the individual qubits.

\subsection{Structural Characterization}
A combination of optical microscopy (OM), scanning electron microscopy (SEM), atomic force microscopy (AFM), and scanning/transmission electron microscopy (S/TEM) was used to obtain structural information from the qubits. OM, SEM, and AFM were used to identify and quantify chemical impurities, processing residue, cracks, and structural defects present in the measured devices. To ensure that these features were not introduced during qubit measurement or the ensuing characterization, the images were compared to optical images taken before qubit measurement. 
The characterization group also measured and tracked the dimensions of various features associated with the device. This included the length, width, and thickness of base metal electrodes (i.e. the shunting capacitor pads) and the aluminum metal Josephson Junction electrodes. Using AFM, we calculated the surface roughness of the base metal in these devices at room temperature and at cryogenic temperatures. Through S/TEM, we also tracked nanoscopic structural information associated with these devices, such as the thickness of any surface oxides, the thickness of the Josephson Junction barrier dielectric and/or capping layers present on metal electrodes, and the dimensions of uncapped regions. This also included imaging the sidewalls of the devices to monitor the angle and shape of the patterned features as well as the etched substrate trench depth. TEM cross sections were prepared from the same area for each qubit within a chip and multiple cross-sections were prepared to determine the spread in these nanoscale features within an individual qubit. Similarly, the grain size of the various metal layers was also monitored.
\subsection{Chemical Characterization}
A combination of analytical S/TEM techniques - energy dispersive spectroscopy (S/TEM-EDS) and electron energy loss spectroscopy (S/TEM-EELS) was applied to capture local chemical information from the qubits. This included evaluating the chemical composition of any surface oxides present on the metal electrodes, as well as the Josephson Junction barrier dielectric layer. It also involved analyzing the level of alloying between adjacent metal layers by characterizing intermediate regions. Additionally, time-of-flight secondary ion mass spectrometry (ToF-SIMS) was used to detect the presence of impurities such as carbon, hydrogen, oxygen, chlorine, and fluorine within the metal films and from nearby exposed substrate surfaces.

\subsection{Physical Property Characterization}
In terms of physical properties, a combination of I-V measurements, magneto/optical (MO) imaging, and near-field terahertz measurements were performed on individual qubit devices. By measuring the I-V response at $\sim$20 mK in each device, we were able to measure the normal state resistance ($R_n$), critical current ($I_c$), and the switching current ($I_{sw}$) in each device using methods discussed previously \cite{Wisne_2024}. With MO imaging, we can visualize the magnetic flux penetration within the superconducting base metals after cooling in a zero magnetic field at 4K. Specifically, we tracked the total area of the region penetrated by magnetic flux as well as the nonuniformity of the flux front. Both parameters are calculated relative to the geometry of the pad itself, thus making them size- and geometry-independent. Additionally, we used terahertz (THz) nanoimaging and nanospectroscopy as a non-destructive tool to probe qubit devices spanning from Nb in its normal state at high temperatures to its superconducting state at cryogenic temperatures below 9K. This emerging technique is empowered by the recent demonstration of an extreme THz scanning near-field optical microscopy (SNOM), operating at temperatures as low as 1.8 K, in magnetic fields up to 5 T, and across a frequency range of of up to 2 THz \cite{10.1063/5.0130680}. In this case, we combine broadband THz pulses with atomic force microscopy to capture variations in the local conductivity with nanoscale resolution. These near-field signals $\textbf{s}_n$ were extracted from the scattered terahertz signal and provided insight into the charge accumulation at edges of the superconducting film as well as the dielectric response of the exposed substrate. 

Because the Nb resonator films used here are approximately 150–200 nm thick, they effectively function as bulk-like high reflectors for the THz light. Consequently, in THz SNOM measurements, any differences between the Nb in its normal and superconducting states appear relatively subtle compared to the pronounced signals generated by inhomogeneities at sharp boundaries, edges, and other nanoscale disorders, which concentrate the electric field \cite{RN45}. The THz near-field microscopy experiments are based on a ytterbium laser with a pulse energy of 20 $\mu$J, a repetition rate of 1 MHz, a pulse width of 91 fs, and a central laser wavelength of 1030 nm. A representative THz near-field image of an Nb film with a nominal thickness of 175 nm at 5.1 K (below its critical temperature) is shown in the supplementary material (Figure \ref{fig:STHz}), along with the scattered THz pulses measured at a single point and their corresponding spectrum. While THz near-field microscopy and AFM can achieve similar spatial resolution and serve as non-destructive techniques for examining trench depth, a key advantage of THz near-field microscopy is its ability to reveal not only geometric steps but also localized and concentrated electric fields around sharp edges, sidewalls, and substrate trenches.
\section{Results}
In the ensuing sections, we focus primarily on the variety of structural, chemical, and physical differences from qubit-to-qubit within a single chip using these techniques as summarized in Figure \ref{fig:figure-summary}. Upon discussing these observations, we then focus on the few parameters that appeared to correlate strongly with qubit-to-qubit variations in the average quality factor and include electromagnetic (EM) simulations as needed to probe the origin of these trends. The quality factor was selected as the primary performance metric of interest to account for the spread in frequency from qubit-to-qubit on a single chip. We finally discuss potential proxy techniques for predicting qubit performance, as well as observed variations from device to device that do not exhibit a trend with performance.
\subsection{Findings}
\begin{figure*}
    {\includegraphics[width =7in]{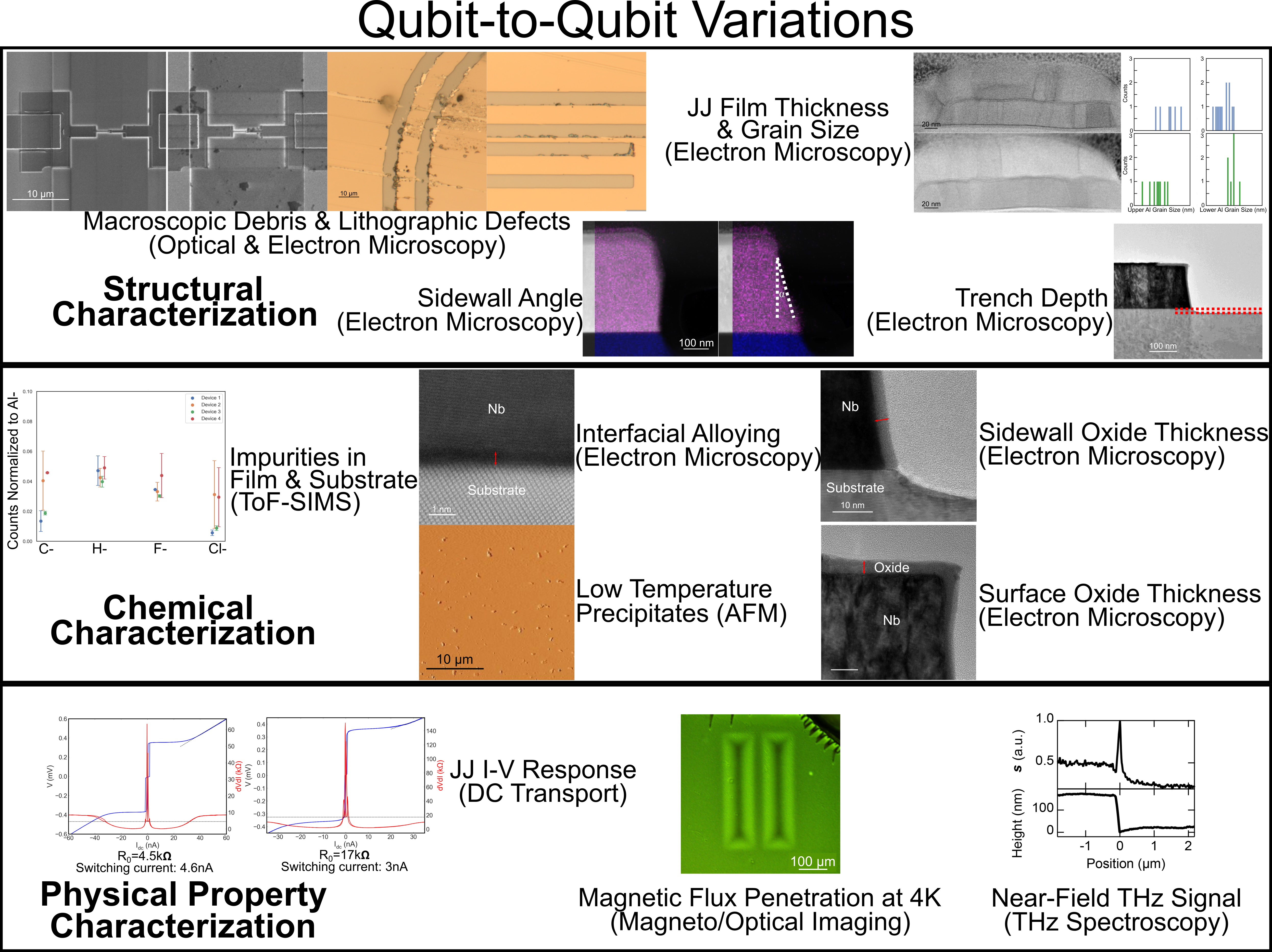}}
   \caption{A summary of the structural, chemical, and physical property differences from qubit-to-qubit within a single chip that were identified in this study. The technique through which each feature or property was measured is provided in the parenthetical.}
   \label{fig:figure-summary}
\end{figure*}

As captured in Figure \ref{fig:figure-summary}, the aforementioned characterization techniques reveal a bevy of structural, chemical, and physical differences from qubit-to-qubit within a single chip. OM and SEM analysis reveal the presence of macroscopic debris and lithographic defects in the vicinity of individual qubits as well as the neighboring readout resonators. S/TEM analysis reveals variations in the thickness and grain size of the Josephson junction leads. We do not observe meaningful differences in the chemistry or the thickness of the AlO$_x$ insulating barrier across qubits as analyzed with S/TEM-EELS. With respect to the superconducting Nb pads, this technique also reveals variations in the sidewall geometry, depth of the etched substrate trench, and oxide thicknesses of the top surface and the sidewall. We also do not observe meaningful differences in the chemistry of these oxides across a single chip.

From ToF-SIMS, we observe variations in the carbon, hydrogen, oxygen, chlorine, and fluorine content within the metal films and from nearby exposed substrate surfaces from qubit to qubit. We do not observe meaningful differences in the level of alloying at metal/metal or metal/substrate interfaces across qubits. Low temperature AFM reveals the presence of non-superconducting niobium hydrides at temperature $<$200K \cite{lee2021discovery} and the density of these precipitates varies across qubits. 

In terms of physical property measurements, DC transport measurements reveal variations in the normal-state resistance as well as switching current across the Josephson junction from qubit to qubit. From MO imaging, we observe variations in the total area of the region penetrated by magnetic flux as well as the non-uniformity of the flux front across qubits. And finally, from THz spectroscopy, we observe variations in the peak near-field THz signal measured at the edge of the superconducting pads from qubit-to-qubit. These observations were subsequently compared to the performance metrics of these qubits to identify properties and features in these devices that appear to trend with qubit quality factor.

\subsection{Relationship between Trench Depth and Performance}
\begin{figure*}
    {\includegraphics[width =7in]{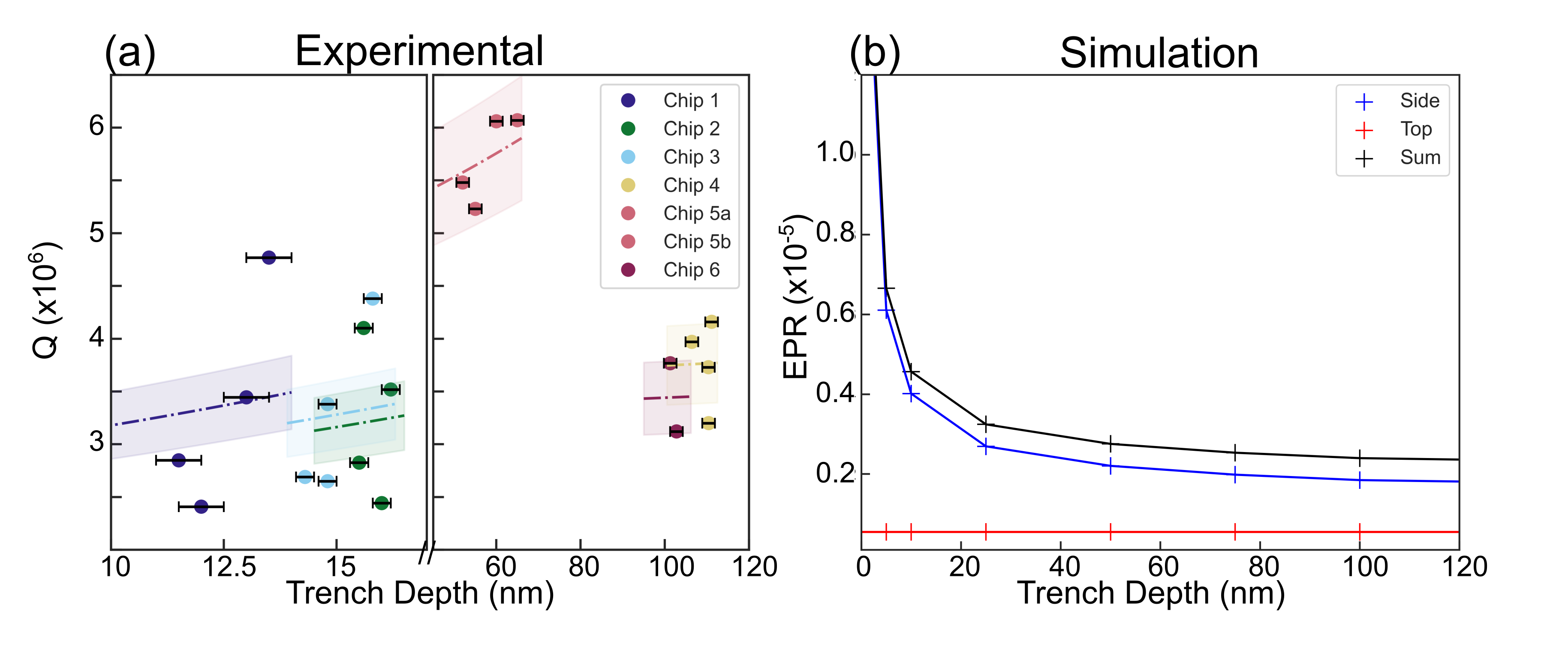}}
   \caption{Effect of variations in trench depth on variations in the quality factor of qubits. (a) Experimental data showing trends between trench depth and device quality factor across different wafers. In Qubits with substrate trench depths less than 20 nm were prepared on sapphire, whereas qubits with larger substrate trench depths were prepared on silicon wafers. (b) Simulated data plotting the surface energy participation ratio (EPR) as a function of trench depth. The shaded regions in (a) represent the level of qubit-to-qubit variation for a single chip expected in the case where trench depth represents the sole source of variation based on the simulated data. This assumes a 10\% error in the loss predicted by the simulations. Given that the performance of several qubits lie outside the shaded regions, it appears that other sources of variation contribute to the variations in performance observed.}
   \label{fig:figure2}
\end{figure*}

In agreement with previous reports by \citeauthor{7745914} and \citeauthor{MURRAY2021100646}, we observe that there appears to be a relationship between increasing trench depth, which represents the depth of recesses etched into the underlying dielectric substrate immediately adjacent to superconducting metal edges, and increased qubit quality factor within qubits on a single chip in this study (Figure \ref{fig:figure2}) \cite{7745914, MURRAY2021100646, 9106772}. An example of the relationship between trench depth and the quality factor of the qubit is provided in Figure \ref{fig:S2} for superconducting qubits prepared with a niobium base metal capped with tantalum metal \cite{RN1} on a silicon substrate.

\begin{figure*}
    {\includegraphics[width =7in]{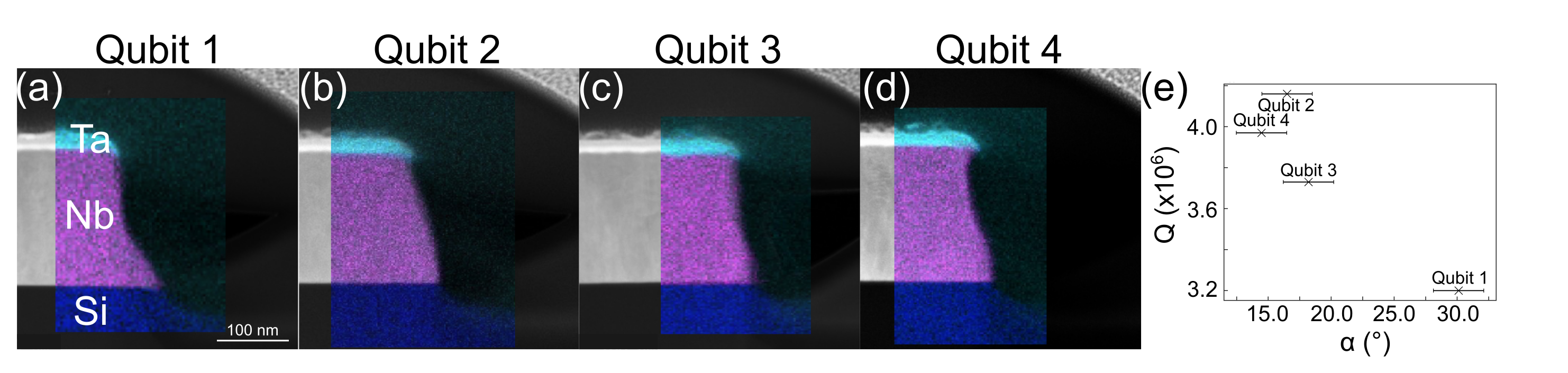}}
   \caption{(a-d) Electron microscopy images taken of the sidewall geometry for qubits 1-4 on Chip 4 consisting of a base metal of Nb film capped with Ta. Energy dispersive spectroscopy (EDS) maps indicating the presence of Ta, Nb, and Si are overlaid. (e) Trend between $\alpha$ as defined in Figure 4 and qubit quality factor for the 4 qubits on this chip. Larger $\alpha$ values appear to trend with decreasing quality factor. The scale bar is consistent for (a-d).}
   \label{fig:figure3}
\end{figure*}

Based on electromagnetic (EM) simulations of the qubit geometry employed in this study, we observe that the energy participation ratio (EPR) of the lossy surface dielectric decreases logarithmically with increases in trench depth and saturates at trench depth values of the order of 70 nm, as seen in Figure \ref{fig:figure2}b. We find that this reduction in the surface EPR primarily arises from a suppression in the field at the sidewalls (side EPR) of the superconducting metallization layer with increasing trench depth as opposed to suppression in the field from the top surface (top EPR). Qualitatively, these simulations agree with the experimental results regardless of the superconducting materials and substrates used, where, similarly, small variations in the trench depth trend with significant variations in the qubit quality factor for qubits with trench depth values less than 20 nm. With increasing trench depth, variations in trench depth appear to induce smaller variations in the qubit quality factor, and in the case where trench depths exceed 70 nm, trench depth appears to be uncorrelated with the qubit quality factor.  However, these variations in trench depth do not entirely explain the qubit-to-qubit variations in performance observed in this study.
Based on the simulated data, the shaded regions in \ref{fig:figure2}a represent the level of qubit-to-qubit variation for a single chip expected in the case where trench depth represents the sole source of variation. This assumes a 10\% error in the loss predicted by the simulations. Given that the performance of several qubits lies outside the shaded regions, it appears that other sources of variation contribute to the variations in performance observed. In the next sections, we will discuss two additional sources that correlate with performance. These sources are most apparent for qubits with substrate trench depths exceeding 70 nm (Chips 4 and 6) and qubits with a top surface consisting of niobium oxide (Chip 2).

\begin{figure*}
    {\includegraphics[width =7in]{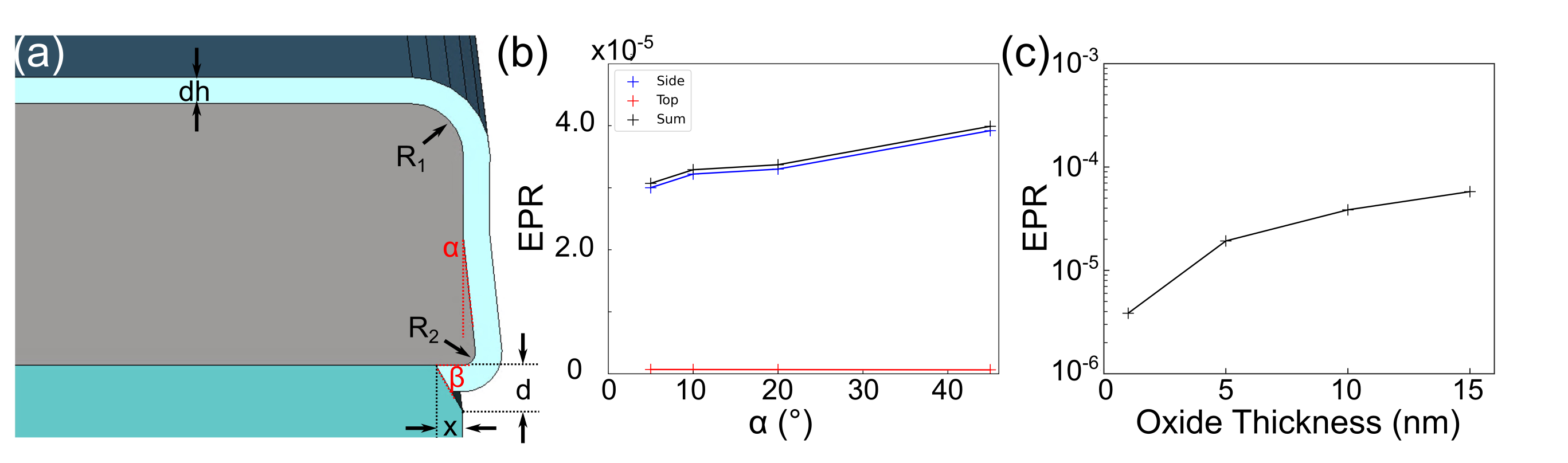}}
   \caption{EM simulations of the qubit sidewall. The sidewall angle ($\alpha$), oxide thickness (dh), radius of curvature of the top edge (R$_1$), radius of curvature of the bottom edge (R$_2$), distance of undercut ($x$), and undercut angle ($\beta$) are defined in (a).  Surface EPR values are provided as a function of (b) $\alpha$ for dh = 5 nm and d = 100 nm as well as (c) dh for $\alpha$ = 0 degrees and d = 100 nm. Top represents the EPR associated with the top surface, side represents the EPR associated with the sidewall surface, and sum represents the total surface EPR, i.e. the sum of these two quantities.}
   \label{fig:figure4}
\end{figure*}

\subsection{Sidewall Geometry}
In the case of qubit devices with large trench depths, we observe a relationship between the sidewall geometry and the qubit quality factor. We observed consistency in the sidewall geometry along the perimeter of each qubit pad, and representative qubit sidewall geometries captured from devices with large trench depths are provided in Figure \ref{fig:figure3} and Figure \ref{fig:S4}. From these images, we observed variations in the top edge curvature, bottom edge curvature, and sidewall footer angle. We hypothesize that these non-uniformities in the sidewall geometry likely arise because the etching process exhibits some level of isotropic character and requiring further optimization. Of these three features, only variations in the sidewall footer angle appear to correlate with the qubit quality factor. Specifically, as seen in Figure \ref{fig:figure3}e, we observe that for multiple chips, increases in this angle (i.e. situations where we observe a smoother decline from the top surface of the metal to the substrate surface) trend with a reduction in the qubit quality factor.

To gain greater insight into the relationship between sidewall footer angle and qubit quality factor, as well as the lack of relationship between other geometrical features and performance, we parameterized the sidewall and performed a series of EM simulations to understand the types of structural variations that have the most significant impact on performance. We varied the sidewall footer angle, represented by $\alpha$ as defined in Figure \ref{fig:figure4}. We also varied the radius of curvature (R$_1$) for both the top surface of the superconducting film and the bottom edge (R$_2$) in Figure \ref{fig:S4}.

As seen in Figure \ref{fig:figure4}, we find a relationship between variations in $\alpha$ and the surface EPR of the sidewall. The simulations indicate that devices with $\alpha$ of approximately 30 degrees (as seen in Figure \ref{fig:figure3}) exhibit surface EPR values that are 15-20\% larger than a comparable device with $\alpha$ of approximately 10-15 degrees. On the other hand, we find that variations in the radius of curvature of the top edge have virtually no impact on the surface EPR due to competing effects between fields in the top surface and sidewall. Similarly, virtually no impact on surface EPR is observed when the radius of curvature of the base of the film varies, which also supports our experimental findings (Figure \ref{fig:S4}).

\subsection{Niobium oxide}
In addition to structural variations, we also observe a trend between surface chemistry and qubit performance. This is specifically observed in the case of Chip 2, which does not display the trend between trench depth and qubit performance that is expected based on simulations. Previous measurements by SQMS researchers as well as groups around the world using 2D resonators, 3D cavities, and qubits have unambiguously demonstrated that the surface oxide that spontaneously forms on the surface of niobium in ambient conditions represents the major source of microwave loss at low powers and mK temperatures \cite{PhysRevApplied.13.034032, PhysRevLett.119.264801, RN33, Niepce_2020,  PhysRevApplied.22.024035, Burnett_2016, RN11, RN14}. Specifically, 3D cavity measurements have shown that the loss tangent of this 5 nm thick oxide is $\sim$~0.1. This value is orders of magnitude larger than the loss tangents calculated at the metal/substrate interface as well as those calculated for the underlying substrate \cite{doi:10.1063/5.0017378, PhysRevApplied.18.034013}. As a result, in addition to the removal of this oxide that leads to a 50-200$\times$ increase in the photon lifetime of 3D Nb SRF cavities in the TLS-dominated ($<$1K) regime \cite{PhysRevApplied.16.014018}, mitigating the formation of this lossy surface through surface encapsulation has led to a 3-5$\times$ boost in measured qubit T$_1$ times \cite{RN1}. Given this context, it is reasonable to predict that variations in surface oxide thickness can affect qubit performance, and EM simulations indicate that the surface EPR can vary from 2.8 $\times 10^{-5}$ to 5.8 $\times 10^{-5}$ when the surface oxide thickness varies from 5 nm to 10 nm, as seen in Figure \ref{fig:figure4}c.

\begin{figure*}
    {\includegraphics[width =7in]{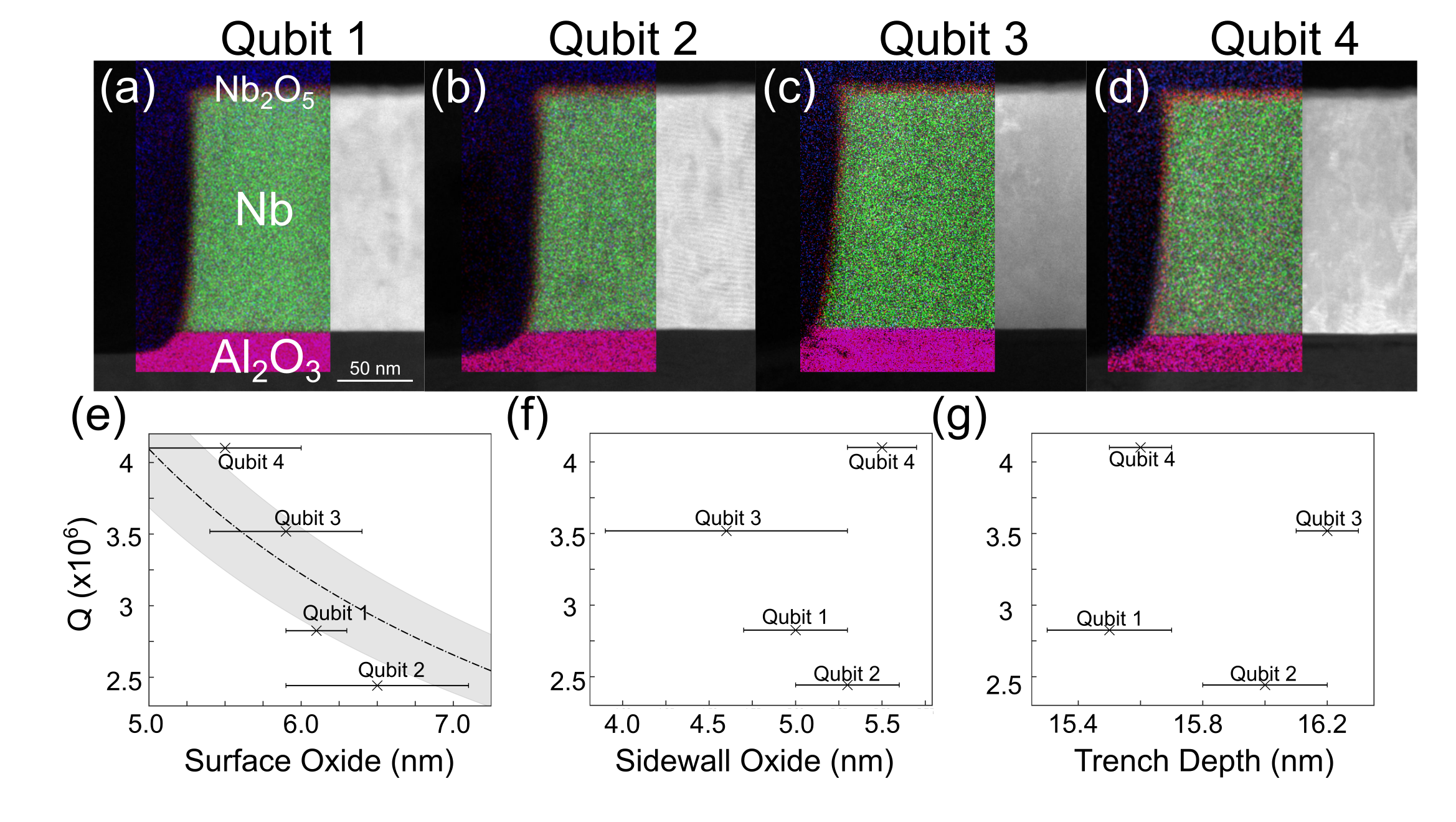}}
   \caption{(a-d) Electron microscopy images taken of the sidewall geometry for qubits 1-4 on Chip 2 consisting of a base metal of Nb film. Nb and O EDS maps are provided as an overlay. Relationships between qubit quality factor and (e) surface oxide thickness, (f) sidewall oxide thickness, and (g) trench depth for the 4 qubits on this chip are provided. Larger surface oxide thickness values appear to trend with decreasing quality factor. The shaded regions in (e) represent the level of qubit-to-qubit variation expected in the case where niobium oxide thickness represents the sole source of variation based on the simulated data. The scale bar is consistent for (a-d).}
   \label{fig:figure5}
\end{figure*}

In particular, the data set associated with Chip 2 in Figure \ref{fig:figure2} represents a set of 4 qubits where the base metal consisted of bare niobium. As seen in Figure \ref{fig:figure5}, although there is no correlation between trench depth and variations in qubit quality factor between devices, increases in niobium oxide thickness, specifically at the top surface, trend with decreases in quality factor. This agrees with the simulations in Figure \ref{fig:figure4}. The bars in the scatter plot represent the range in thicknesses observed for each device and indicate that the surface oxide thickness of this polycrystalline film of niobium varies from 5 nm to 7 nm across a single 7.5 mm by 7.5 mm chip. The shaded regions in \ref{fig:figure5}e represent the level of qubit-to-qubit variation expected in the case where the thickness of the niobium oxide represents the sole source of variation based on the simulated data.

\begin{figure*}
    {\includegraphics[width =7in]{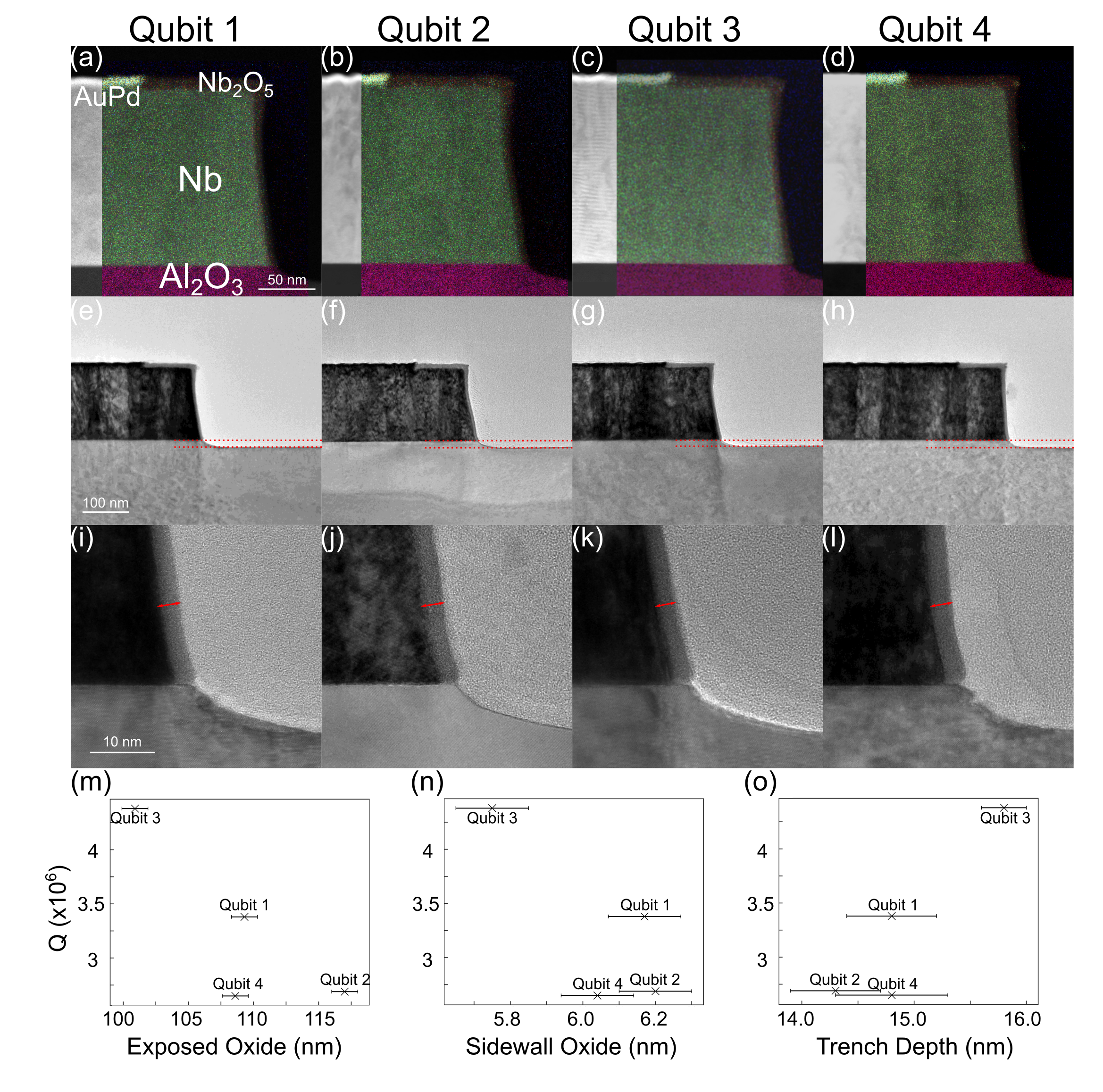}}
   \caption{(a-d) Electron microscopy images taken of the sidewall geometry for qubits 1-4 on Chip 3, consisting of a base metal of Nb capped with AuPd metal. Nb O, and Au EDS maps are provided as an overlay. Electron microscopy images of the sidewall, with the (e-h) trench depths and (i-l) sidewall oxide indicated in red. Relationships between qubit quality factor and (m) length of exposed oxide, (n) sidewall oxide thickness, and (o) trench depth for the 4 qubits on this chip are provided. In addition to a trend between trench depth and increasing qubit quality factor, we observe a weak trend between exposed oxide regions and decreasing quality factor. The provided scale bars are consistent across each row.}
   \label{fig:figure6}
\end{figure*}

The impact of this niobium oxide on variations in qubit quality factor is also apparent from a 4 qubit data set composed of niobium capped with AuPd metal as the base metal (Chip 3). In Figure \ref{fig:figure6}, we observe that for this set of devices, the capping layer does not completely protect the top surface of the niobium. Rather, a region extending roughly 100 nm from the sidewall is left exposed and oxidizes. We suspect that this was a result of isotropic etching introduced by the gold etchant solution before dry anisotropic etching of niobium, as discussed previously \cite{RN1}. The length of this exposed niobium region presents a new variable in the case of these devices and varies from qubit to qubit. This is likely a result of local variations in the quantity of etchant across the wafer. Notably, no meaningful differences were observed in the oxide chemistry from device to device for the case of these samples. In terms of the impact on performance, we observe a trend between increases in this length of exposed niobium and a decrease in the quality factor, in addition to a trend between increasing trench depth and increasing quality factor. As such, these two chips indicate that in the case where niobium oxide persists at the top surface, variations in the oxide thickness may have a similar or more significant role on performance variations across neighboring qubit devices compared to trench depth. 

\subsection{Proxy Techniques}
The slow turnaround times associated with qubit fabrication and mK device characterization currently limit researchers' ability to test and verify the impact of new surface or material processing strategies on device performance \cite{deLeoneabb2823, doi:10.1063/5.0017378}. Although rapid turnaround cryogenic systems represent a critical piece in addressing this challenge, developing proxy techniques that can be used to assess samples after each fabrication step in the context of achievable qubit performance is needed for improving device reliability. To this end, we have identified two proxy techniques through this study.

\begin{figure}
    {\includegraphics[width =\columnwidth]{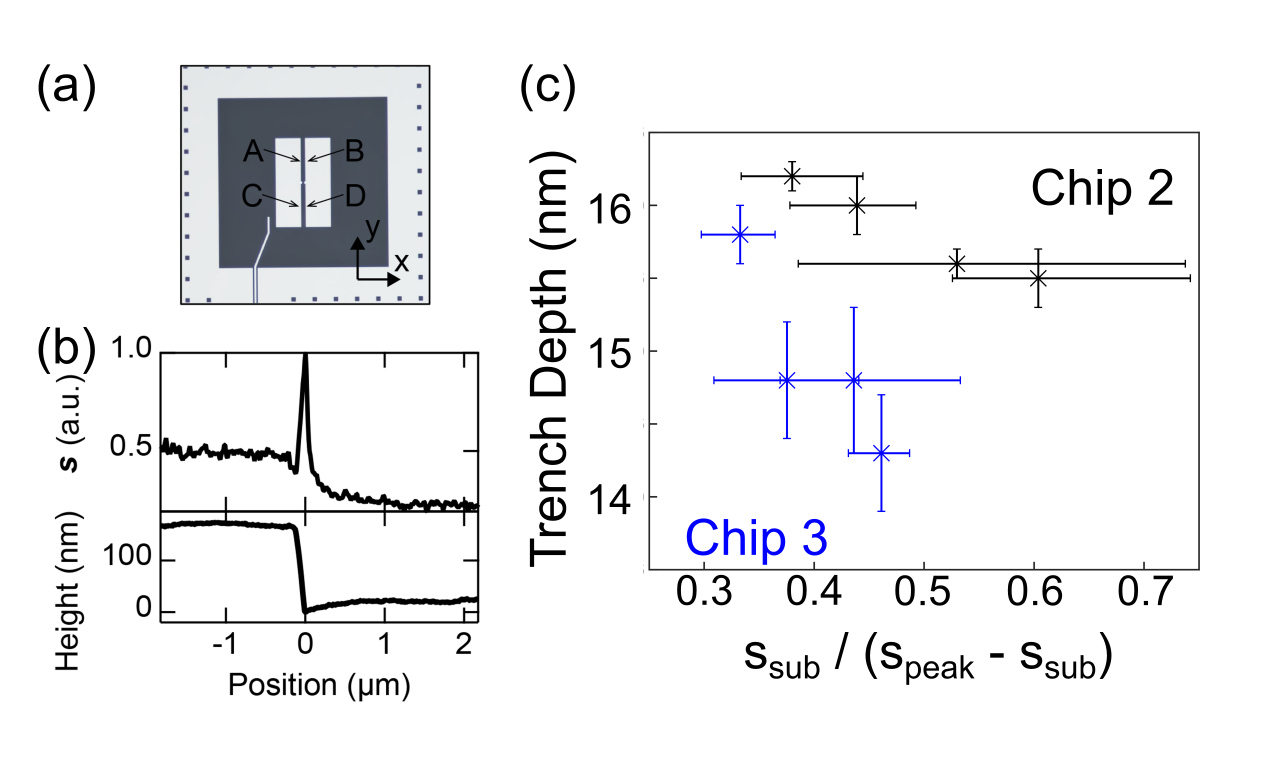}}
   \caption{Near-field THz spectroscopy data taken from Chips 2 and 3. (a) Scans were performed at 4 locations on each device (regions marked A, B, C, and D). An example of near-field signal and an AFM height profile taken from the edge of a superconducting pad are provided in (b). As seen in (c), peak near-field intensity normalized by the near-field intensity measured at the substrate scales with decreasing trench depth for both qubits with base layers consisting of niobium and base layers consisting of niobium capped with a metal.}
   \label{fig:figure7}
\end{figure}

Given the relationship between trench depth and qubit quality factor we have observed, developing a non-destructive characterization method for identifying sub-nm scale variations in trench depth from qubit to qubit at early stages of the fabrication process may offer opportunities to improve device uniformity through corrective actions or pre-selection of homogeneous areas of the wafer. Due to its destructive nature, S/TEM characterization of metallization edges following FIB liftout is not a suitable method for such an evaluation. Instead, we find that the near-field signal as measured by THz spectroscopy offers an opportunity in this context, as exhibited in Figure \ref{fig:figure7} \cite{RN45}. By normalizing the peak near-field intensity by the near-field intensity measured at the substrate, we observe that this metric scales with decreasing trench depth for both qubits with base layers consisting of niobium and base layers consisting of niobium capped with a metal. Simulations are currently underway to understand the origins of this relationship, and these results will be discussed more extensively in an upcoming publication.

\begin{figure}
    {\includegraphics[width =\columnwidth]{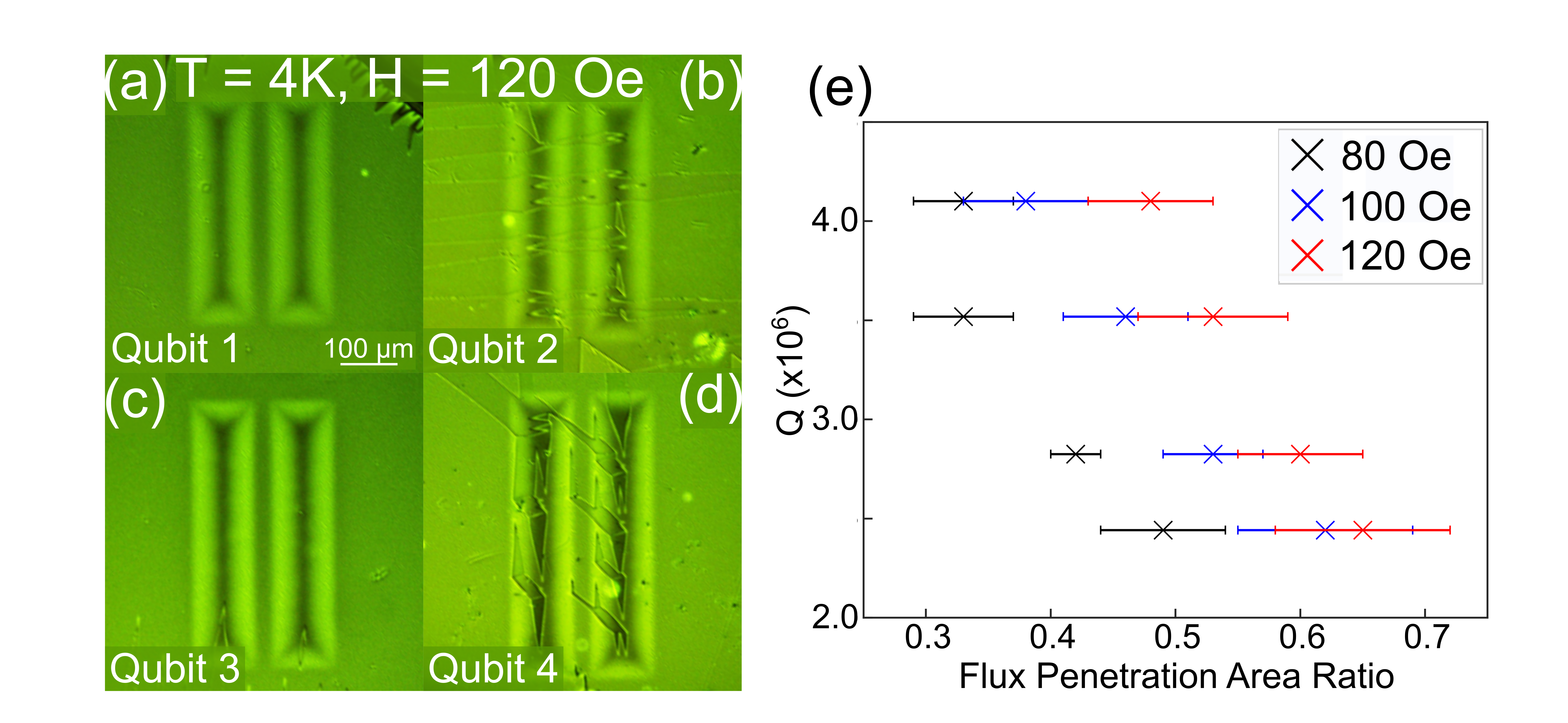}}
   \caption{(a) MO imaging data taken from Chip 2 following cooling in zero magnetic field. (b) The total area of the region penetrated by magnetic flux in superconducting niobium measured is plotted against the qubit quality factor for external magnetic fields of 80, 100, and 120 Oe. We observe that this metric scales with decreasing qubit quality factor.}
   \label{fig:figure8}
\end{figure}

Moreover, initial MO imaging results at 4K suggest that monitoring magnetic flux penetration offers an opportunity to both predict variations in performance across a set of qubits on a single chip and observe non-uniformities that are relevant to performance within individual qubits. The former trend is illustrated in Figure \ref{fig:figure8}. In this case, following cooling in zero magnetic field, the total area of magnetic flux penetration in superconducting niobium scales inversely with the qubit quality factor. Similar trends are observed for external magnetic fields of 80, 100, and 120 Oe. The origin of this relationship is the subject of further analysis, but based on the S/TEM data presented in Figure \ref{fig:figure6}, we hypothesize that variations in the surface oxide thickness between these qubits may induce variations in the superconducting state between devices.

\begin{figure}
    {\includegraphics[width =\columnwidth]{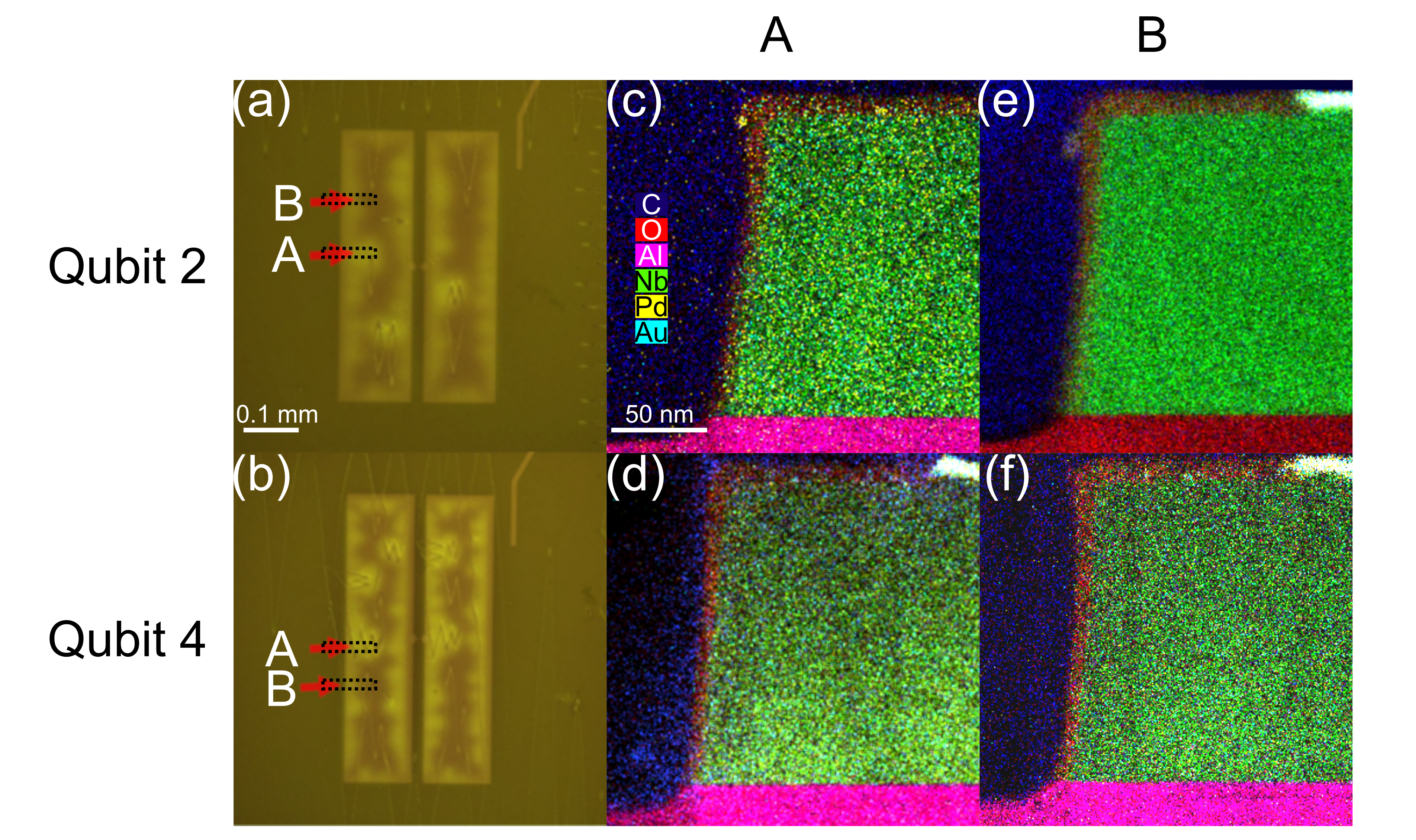}}
   \caption{(a-b) MO imaging data taken from Chip 3 following cooling in zero magnetic field. The magnetic flux penetration is highly non-uniform in this case. (c-d) Using electron microscopy, we find that regions with increased flux penetration preferentially exhibit greater amounts of exposed Nb surface oxide compared to (e-f) regions with decreased magnetic flux penetration. Nb and O EDS maps are provided as an overlay. The scale bar is consistent for (c-f).}
   \label{fig:figure9}
\end{figure}

In the case of a metal-capped niobium sample, the penetrating magnetic flux is highly nonuniform, as seen in Figure \ref{fig:figure9}. By using S/TEM to analyze regions with increased magnetic flux penetration as well as regions with reduced magnetic flux, we observe that regions with increased magnetic flux penetration also exhibit larger regions of exposed Nb top surface that extend from the sidewall compared to regions with decreased magnetic flux penetration. This finding represents a plausible explanation, since such variations in the uniformity of the superconducting state can produce non-uniform magnetic flux fronts. Additional theoretical investigation to decipher the origin of the potential correlation between flux penetration and qubit performance is currently underway. However, these initial proof-of-concept data points suggest that MO imaging may serve as a relatively simple way to test and screen materials and fabrication processes at 4K. 

\subsection{Inconclusive Impact on Performance}
In addition to the aforementioned features that exhibit correlations with qubit performance, we observed variations in several structural, chemical, and physical features from device to device, for which the impact on performance is presently unclear. These sources may help explain some of the performance variations that were not fully explained by variations in trench depth, oxide thickness, and sidewall geometry. Disentangling the impact of these parameters on qubit performance will require future studies systematically varying them and performing A/B tests. These parameters are illustrated in the Supporting Information and are summarized below.
\subsubsection{Macroscopic Debris and Lithographic Defects}
As captured in Figures \ref{fig:S5} and \ref{fig:S6}, macroscopic debris and lithographic defects were observed in several of the measured qubit devices. Based on the chemical nature of the observed debris, it appears that the debris took one of a few forms: (1) metal residue from incomplete lift-off, (2) chemical residue introduced during solvent rinsing, (3) scratches during wafer handling, or (4) particles likely introduced during wirebonding. Meanwhile, lithographic defects such as photoresist residue, as well as reduced pattern fidelity in high curvature regions, were observed across several devices. In both cases, no obvious trends with variations in qubit quality factor were observed. A detailed study involving the systematic placement of debris and lithographic defects is likely needed to isolate and examine the impact of debris as a function of size and location.
\subsubsection{Josephson Junction: Structure and Electrical Properties}
As illustrated in Figures \ref{fig:S7} and \ref{fig:S8} and discussed previously, variations in grain structure and sidewall geometry of the Josephson Junction electrodes, as well as I-V response, are observed for the measured qubit devices. Concerning grain structure, we observe that the median grain size in the upper aluminum lead can vary from 30 nm to 50 nm, depending on the qubit. Similarly, in the lower aluminum lead, we observe that the median grain size varies from 20 nm to 35 nm. In both cases, there does not appear to be any relationship with the qubit quality factor. Similar variations to those observed in the sidewall angle of superconducting pads are observed for the aluminum leads, as captured in Figure \ref{fig:S7}. In this case, however, no clear trends between sidewall angle and qubit performance are observed. Finally, in terms of the I-V response, device-to-device variations are also observed in terms of normal state resistance as well as switching current (Figure \ref{fig:S8}). Once again, no trend is observed between these parameters and qubit performance, which thereby suggests that in the case of the high-coherence superconducting qubits investigated in this study, variations in the Josephson Junction morphology do not appreciably alter the overall performance.
\subsubsection{Surface Chemistry}
The resulting average impurity counts measured with ToF-SIMS are provided in Figure \ref{fig:S9} Across the three analyzed chips, there do not appear to be any obvious trends between the surface chemistry of the superconducting pads and the qubit performance. Potentially, there may be a trend between residual fluorine content from plasma etching processes and performance, however, increased statistics are needed to isolate the impact of this impurity on performance. Moreover, although hydrogen impurities are known to bind with niobium at cryogenic temperatures and form non-superconducting precipitates, as reported by \citeauthor{lee2021discovery} and seen in Figure \ref{fig:S10}, no consistent trend is observed between hydrogen impurity content and performance across all qubits\cite{lee2021discovery}. This is in agreement with recent work suggesting that hydrogen impurities in niobium may not be linked to TLS loss \cite{https://doi.org/10.1002/adfm.202401365}. Additionally, we also probed the chemistry of the substrate surface adjacent to the measured qubits. In this case, the spot-to-spot variation in the substrate surface chemistry for a single qubit was found to be similar to the variation in substrate surface chemistry observed across devices. Thus, dedicated studies where the surface chemistry is locally altered are needed to understand the impact of impurities in superconducting pads and substrate regions on device performance.
\subsubsection{Sidewall Oxide Thickness}
Similar to variations in the niobium top surface oxide thickness observed in Figure \ref{fig:figure5}, variations in the niobium sidewall oxide thickness are also observed and reported. Unlike our observations for the niobium surface oxide, there does not exist an obvious correlation between variations in niobium sidewall oxide thickness and variations in performance. This may be due to two factors. First, unlike the 1-2 nm variations observed in the niobium surface oxide thickness, the variations in the niobium sidewall oxide thickness are of a smaller magnitude and typically range from 0.5 nm -1 nm. Once again, an additional dedicated study focused on varying the thickness of the niobium oxide at the sidewall is needed to meaningfully extract the relative impact variations in this parameter imposed on performance.

\section{Discussion}
Together, these results suggest that obtaining an array of high T$_1$ transmon qubit devices with a low spatial variation in the average T$_1$ times requires optimizing the etching conditions to obtain (1) sufficiently large trench depths, (2) uniform sidewall geometries, and careful film preparation to (3) eliminate the presence of niobium oxide. 

Regarding the first point, our results indicate that device-to-device variations are suppressed when trench depths exceed 70 nm. Although achieving these trench depths is straightforward when silicon is employed as the substrate material, the strong covalent bonds inherent to sapphire make it resistant to chemical reactions present in dry etching processes. As a result, trench depths in sapphire were limited to 15 nm -20 nm in this study. An approach to increasing the depth of the trench would be to significantly increase the etch time. However, this may necessitate the use of a metal hard mask as opposed to a polymer photoresist to ensure mask integrity throughout the etching process. Alternatively, one could initially etch the superconducting layer before subjecting the sample to a deep reactive ion etching or an alternative reaction chemistry to more rapidly and selectively remove sapphire. 

Another option to improve qubit performance would be to explore a more prominent undercut than observed in the qubits examined in this study by partially removing a portion of the substrate beneath the superconducting film. The simulation results illustrate the effect of this type of undercut on the surface EPR are presented in Figure \ref{fig:S11} in agreement with previous studies \cite{PhysRevApplied.12.014012}. In Figure \ref{fig:S11}, $x=0$ represents the situation in which a highly anisotropic etch process is applied, and a trench is prepared in the substrate without producing an undercut underneath the superconducting film. In this plot, larger values of $x$ signify larger undercuts of the substrate. Based on the simulations, we find that an undercut on the order of 60 nm can reduce the surface EPR by 50-75\%. This suggests that developing a separate selective etch recipe for a substrate that is potentially more isotropic may be beneficial. At the same time, we find from the simulation results that the undercut processes must be performed strategically. Specifically, the results suggest that shallow undercut angles, as seen in Figure \ref{fig:S11}, can trap EM fields and instead increase the surface EPR compared to larger undercut angles. 

Concerning sidewall geometries, reproducible sidewalls can be achieved by developing more anisotropic dry etching recipes for the metal. To this end, sidewall imaging with electron microscopy can be a valuable tool for recipe optimization, while parameters such as etching chemistry (F-based vs Cl-based), level of physical bombardment (Ar content), and power are varied. Finally, with respect to surface encapsulation \cite{RN1}, our results indicate that incomplete coverage, especially close to the sidewall, can have a noticeable detrimental effect on device performance. This was the case for niobium films capped with AuPd as opposed to niobium films capped with Ta.

In summary, we conducted a coordinated, blind study aimed at understanding sources of qubit-to-qubit variations in high-coherence transmon qubits. Through a combination of structural, chemical, and physical characterization techniques in conjunction with EM simulations, we observed that trench depth, sidewall geometry, and niobium surface oxide thickness all appear to contribute to variations in the qubit quality factor. Additionally, we provide a list of features that varied from device-to-device, for which the impact on performance remains unclear. Furthermore, we identify two potential low-temperature characterization techniques that may potentially serve as proxy tools for qubit measurements. Together, the knowledge gained from this study provides a pathway not only to reduce performance variations across neighboring devices, but also to engineer and fabricate optimal devices to achieve performance metrics beyond the state-of-the-art values. Finally, this study provides a framework for future blind, coordinated investigations that combine device characterization with structural, chemical, and physical characterization tools to decipher and understand other sources of performance spread, including variations in TLS density or variations in T$_1$ with time from device to device, in superconducting qubits.



\begin{acknowledgments}
This material is based upon work supported by the U.S. Department of Energy, Office of Science, National Quantum Information Science Research Centers, Superconducting Quantum Materials and Systems Center (SQMS) under contract number DE-AC02-07CH11359. This work made use of the Pritzker Nanofabrication Facility of the Institute for Molecular Engineering at the University of Chicago, which receives support from Soft and Hybrid Nanotechnology Experimental (SHyNE) Resource (NSF ECCS-2025633), a node of the National Science Foundation’s National Nanotechnology Coordinated Infrastructure. This work also made use of the EPIC facility of Northwestern University’s NU\textit{ANCE} Center, which has received support from the SHyNE Resource (NSF ECCS-2025633), the International Institute of Nanotechnology (IIN) (NIH-S10OD026871), and Northwestern's MRSEC program (NSF DMR-2308691). Research reported in this publication was supported in part by instrumentation provided by the Office of The Director, National Institutes of Health of the National Institutes of Health under Award Number S10OD026871. The content is solely the responsibility of the authors and does not necessarily represent the official views of the National Institutes of Health. This work also made use of instruments in the Sensitive Instrument Facility in Ames National Laboratory. Ames National Laboratory, including the Sensitive Instruments Facility, is operated for the U.S. DOE by Iowa State University under Contract No. DE-AC02-07CH11358. Parts of this work made use of Rigetti Fab-1 and measurement facilities. NPL acknowledges the UKRI International Science Partnership Fund (ISPF)
\end{acknowledgments}

\bibliography{main.bib}

\begin{table*}[t] \label{qubit-results}
\begin{tabular}{|c|>{\centering\arraybackslash}p{2cm}|>{\centering\arraybackslash}p{1.5cm}|>{\centering\arraybackslash}p{1.5cm}|>{\centering\arraybackslash}p{2cm}|>{\centering\arraybackslash}p{2cm}|>{\centering\arraybackslash}p{2cm}|>{\centering\arraybackslash}p{2cm}|}
\hline
     \textbf{Chip ID} &\textbf{Substrate}& \textbf{Nb Film Thickness}& \textbf{Capping Layer}& \textbf{Qubit 1}& \textbf{Qubit 2}&\textbf{Qubit 3}& \textbf{Qubit 4}\\
\hline
     1& c-plane HEMEX Sapphire, Unannealed & 150 nm& 9 nm Au & T$_1^{avg}$=116$\mu$s \newline  $\omega _q$=4.726 GHz \newline Q = $3.44\times10^6$ & T$_1^{avg}$=94$\mu$s \newline  $\omega _q$=4.078 GHz \newline Q = $2.41\times10^6$ & T$_1^{avg}$=109$\mu$s \newline  $\omega _q$=4.158 GHz \newline Q = $2.85\times10^6$ & T$_1^{avg}$=181$\mu$s \newline  $\omega _q$=4.193 GHz \newline Q = $4.77\times10^6$\\
\hline
     2& c-plane HEMEX Sapphire, Unannealed & 163 nm& -- & T$_1^{avg}$=87$\mu$s \newline  $\omega _q$=5.169 GHz \newline Q = $2.83\times10^6$ & T$_1^{avg}$=77$\mu$s \newline  $\omega _q$=5.048 GHz \newline Q = $2.44\times10^6$ & T$_1^{avg}$=120$\mu$s \newline  $\omega _q$=4.667 GHz \newline Q = $3.52\times10^6$ & T$_1^{avg}$=139$\mu$s \newline  $\omega _q$=4.696 GHz \newline Q = $4.10\times10^6$\\
\hline
     3& c-plane HEMEX Sapphire, Unannealed & 155 nm& 6 nm AuPd & T$_1^{avg}$=112$\mu$s \newline  $\omega _q$=4.798 GHz \newline Q = $3.38\times10^6$ & T$_1^{avg}$=91$\mu$s \newline  $\omega _q$=4.697 GHz \newline Q = $2.69\times10^6$ & T$_1^{avg}$=154$\mu$s \newline  $\omega _q$=4.530 GHz \newline Q = $4.38\times10^6$ & T$_1^{avg}$=96$\mu$s \newline  $\omega _q$=4.387 GHz \newline Q = $2.65\times10^6$\\
\hline

     4& Double side polished Silicon & 175 nm & 10 nm Ta & T$_1^{avg}$=101$\mu$s \newline  $\omega _q$=5.039 GHz \newline Q = $3.20\times10^6$ & T$_1^{avg}$=143$\mu$s \newline  $\omega _q$=4.635 GHz \newline Q = $4.16\times10^6$ & T$_1^{avg}$=127$\mu$s \newline  $\omega _q$=4.672 GHz \newline Q = $3.73\times10^6$ & T$_1^{avg}$=144$\mu$s \newline  $\omega _q$=4.391 GHz \newline Q = $3.97\times10^6$\\
\hline
     5a& Double side polished Silicon & 175 nm & 10 nm Ta & & & T$_1^{avg}$=208$\mu$s \newline  $\omega _q$=4.635 GHz \newline Q = $6.06\times10^6$ & T$_1^{avg}$=216$\mu$s \newline  $\omega _q$=4.471 GHz \newline Q = $6.07\times10^6$ \\
\hline
     5b& Double side polished Silicon & 175 nm & 10 nm Ta & & & T$_1^{avg}$=184$\mu$s \newline  $\omega _q$=4.740 GHz \newline Q = $5.48\times10^6$ & T$_1^{avg}$=184$\mu$s \newline  $\omega _q$=4.520 GHz \newline Q = $5.23\times10^6$ \\
\hline
     6& Double side polished Silicon & 170 nm & -- & & & T$_1^{avg}$=108$\mu$s \newline  $\omega _q$=4.600 GHz \newline Q = $3.12\times10^6$ & T$_1^{avg}$=138$\mu$s \newline  $\omega _q$=4.345 GHz \newline Q = $3.77\times10^6$ \\

\hline
\end{tabular}
\caption{\label{tab:fits} Measured Qubit Values. As part of the blind study, the qubit results were not revealed to the characterization team until after the materials and superconducting characterization measurements were completed. Note: Chips 5a and 5b were taken from the same wafer and measured at the same facility.}
\end{table*}
\newpage

\clearpage
\section*{Author Contributions}
\noindent The project was conceived by A.A.M., A.R., and A.G. The devices were fabricated by M.B., F.C., S.G., H.C., J.M., D.O., and P.H. The performance of the devices was measured by S.Z. and E.O.L. Characterization of the samples was performed and analyzed by A.A.M., M.J.B., R.K.C., V.C., A.D., Y.D., C.D.M.D., V.P.D., D.A.G.-W., S.G., D.P.G., S.d.G., S.H., M.C.H., D.I., K.J., R.K., S.K., C.J.K., M.J.K., E.O.L., J.L., P.G.L., A.L., W.M., J.Y.M., J.O., D.P.P., J.P., R.P., R.d.R., D.S., Z.S., M.T., M.J.W., J.W., H.W., and L.W. A.M. wrote the manuscript. All authors discussed the results and the manuscript.

\section*{Competing Interests}
\noindent All authors declare no financial or non-financial competing interests. 

\section*{Supplementary Information}
\noindent See supplementary information for wiring diagrams, pulse schemes associated with device measurements, chemical maps, depth profiles and electron diffraction patterns taken from the samples.

\section*{Disclaimer}
Certain commercial equipment, instruments, or materials are identified in this paper in order to specify the experimental procedure adequately. Such identification is not intended to imply recommendation or endorsement by NIST, nor is it intended to imply that the materials or equipment identified are necessarily the best available for the purpose.

\mbox{~}
\clearpage
\newpage
\setcounter{figure}{0}
\setcounter{page}{1}
\setcounter{section}{0}
\setcounter{table}{0}

\renewcommand{\thepage}{S\arabic{page}}
\renewcommand{\thesection}{S\arabic{section}}
\renewcommand{\thetable}{S\arabic{table}}
\renewcommand{\thefigure}{S\arabic{figure}}

\section*{Supplementary Information}
\section*{Supplementary Methods} \label{Methods}
\subsection{Qubit Fabrication}
22 qubits on 7 chips were fabricated between Pritzker Nanofabrication facility (PNF) and Rigetti Computing's Fab-1 facility. In terms of base metal, these include niobium thin films on the order of 200 nm as well as surface encapsulated niobium with metals including Ta, Au, and AuPd. In terms of the substrate, qubits were prepared on double-side polished, HEMEX grade HEM Sapphire at PNF and were prepared on double-side polished, high-$\rho$ Si substrates (slightly
n-type, \> 10,000 $\Omega \cdot $cm) at Fab-1. More details regarding the film deposition, lithography, etching and JJ fabrication parameters is provided in \cite{RN1}.
\subsection{EM simulations}
The transmon model with a lumped inductance as the Josephson junction and perfect electrical boundary conditions is simulated using ANSYS HFSS code to find the qubit eigenfrequency. The sidewalls of the qubit pad are parameterized and parameters corresponding to the sidewall angle, radius of curvature, undercut, and oxide thickness are varied. Since the typical transmon operating frequency is around 5 GHz and its wavelengths are much larger than the nanometer scale of the sidewall rounding, the HFSS simulation is close to the electrostatic regime, where the exponential behavior of the surface charge density with distance from the edge can be estimated analytically for an arbitrary opening angle. We compared the calculated electric field with the electrostatic approximation for a right angle, which predicts an inverse dependence on distance, and found good agreement. 

We calculate the energy participation ratio ($EPR$), using the following relation. $ EPR_{i} = \frac{\frac{1}{2}\epsilon_0 \epsilon^{'}_i \int |E|^2dV_i}{W_0}$. Here, $i$ represents the number of lossy layers, $EPR_{i}$ represents the EPR in the $i^{th}$ layer, $\epsilon$ represents the complex permittivity of the media, $W_0$ represents the total stored energy, $E$ represents the electric field strength, and $V$ represents the volume.

It is worth mentioning that the fields in the HFSS eigenmode solutions are by default normalized to 1 J of stored energy, but the normalization is only performed for fields in the volumes of physical objects and does not consider the energy stored in lumped elements. Thus, in this case, it is necessary to explicitly calculate the stored energy for the correct normalization of the fields. Finally, an important step to ensure the accuracy of the EPR calculations is a convergence study. The HFSS eigenmode solver uses the finite element method (FEM) to approximate the electromagnetic fields in the simulation domain. The program allows to specify the mesh elements of the 1st or 2nd order, corresponding to a linear and quadratic approximation of electromagnetic fields within each element. We ran a series of simulations, varying the order of approximation and gradually increasing the number of mesh elements. The initial mesh is seeded to ensure a denser mesh in areas of high electric field, and then the simulation goes through three iterative steps before recording the EPR value. We then performed all subsequent EPR analysis using the optimal mesh setting.

\subsection{Materials Characterization}

\subsubsection{Electron Microscopy}
For chips 2 and 3, secondary electron images for the qubit surfaces were acquired on SEM (Helios, Thermo Fisher Scientific Ltd.). 

Cross-sectional TEM samples were prepared by focused ion beam (FIB) using the FEI Helios Nanolab 600 Dual Beam FIB/SEM. Bulk-out and lift-out processes were performed using a 30kV Ga+ ion beam, while final cleaning steps were performed at 5kV and 2kV to remove amorphous materials at the surface. The final thicknesses of the samples were approximately 50-100 nm. For chips 2 and 3, the TEM samples were investigated using an aberration-corrected TEM (Titan Cube, Thermo Fisher Scientific Ltd.) with Super-X EDS detector at 200 kV. For chips 1, 4, 5a, 5b, and 6, TEM and S/TEM data were collected on an aberration-corrected JEOL ARM200CF S/TEM operating at 200kV with a convergence angle of 25 mrad for ADF-STEM imaging in the case of chips 4 and 6 and  a convergence angle of 21 mrad for ADF-STEM imaging in the case of chips 1, 5a, and 5b. EDS data were collected in the same microscope, also at 200kV, with a Dual SSD EDS detector (1.7 steradians). Dual electron energy loss spectra (EELS) were acquired on a JEOL ARM300F GrandARM S/TEM operating at 300kV using a Gatan GIF Quantum energy filter on a K3 IS pixelated detector. Data processing (signal mapping, background subtraction, denoising, plural scattering removal, etc.) was conducted using the Gatan Microscopy Suite (GMS) software.

\subsubsection{THz Spectroscopy}
The terahertz near-field microscopy experiments are based on a ytterbium laser with a pulse energy of 20 $\mu$J, a repetition rate of 1 MHz, a pulse width of 91 fs, and a central laser wavelength of 1030 nm. Ultrafast terahertz pulses are emitted from an organic crystal, and the terahertz near-field signals are detected by electro-optical sampling. The atomic force microscopy (AFM) incorporated here is a cantilever-based tapping-mode system designed for applications at low temperature and in high magnetic fields. The metallic AFM probe is a Rocky Mountain Nanotechnology, LLC model 25Pt300B that has an 80 $\mu$m long platinum coated probe with an operating frequency of 20 kHz (±30\%). A nominal tip radius of 20 nm determines the spatial resolution of the near-field images. Once the tip is in contact with the sample, a time-domain terahertz spectroscopy can be performed by moving the motorized stage that controls the time delay of the optical sampling pulse to the electro-optic crystal and thus trace out the oscillating electric-field waveform of the scattered terahertz near-field amplitude. To obtain near-field images, the sample stage underneath the tip was raster scanned, while the terahertz sampling delay is fixed to a position that gives the largest near-field signal amplitude. The AFM height information is also taken simultaneously along with the near-field image scan. Near-field signals $\textbf{s}_n$ are extracted from the scattered terahertz signal by demodulating the backscattered radiation collected from the tip–sample system at nth harmonics of the tip-tapping frequency from the AFM (n = 1, 2, 3, and 4). Raster scans for near-field imaging are performed with a sampling time of 90 ms per pixel and a pixel size of 10 nm and 20 nm for the Nb sidewall area and the Al Josephson Junction area, respectively."

\subsubsection{Magneto/optical imaging}
Magneto-optical (MO) imaging involves the principle of Faraday rotation to visual magnetic induction on the sample surface with the help of an indicator film on the sample surface. This indicator film is an optically transparent ferrimagnetic film composed of bismuth-doped iron garnet deposited on a transparent substrate. A mirror layer is deposited over the film. During the measurement, the indicator, with its reflective side facing down, is placed on the flat surface of the sample being studied.  Linearly polarized light passes through the substrate and garnet film and then reflects back to the objective without exposing the sample itself. Consequently, the images captured only reveal magnetic fields beneath the indicator. A low-temperature closed-cycle optical cryo-station, Model-s50, from Montana Instruments and the Olympus BX3M optical polarized-light microscope with long focal length objectives is used for magneto-optical imaging. The setup consists of a gold-plated cold post inside the vacuum chamber of the cryostat, with the sample mounted on top. The sample temperature is measured by a Cernox-1030 thermometer located in close proximity to the sample and is controlled from the 3.8 K up to room temperature. A copper-wound solenoid is used to apply an external magnetic field along the light propagation direction. More details of this technique and its applications is provided in Ref. \cite{ChJooss_2002, Datta_2024}.

\subsubsection{Time of flight secondary ion mass spectrometry (ToF-SIMS)}
ToF-SIMS measurements were taken from qubit devices and neighboring substrate regions using a dual beam time-of-flight secondary ion mass spectrometry (IONTOF 5) to analyze the concentration and depth distribution of impurities. Secondary ion measurements were performed using a liquid bismuth ion beam (Bi$^+$). A cesium ion gun with an energy of 500 eV was used for sputtering the surface for depth profile measurements to enhance detection of anions \cite{Bose_2015}. A 25$\mu$m by 25$\mu$m analysis area and 100$\mu$m by 100$\mu$m sputter area was used for all measurements.

\subsubsection{I-V measurements}
Low-temperature dc transport measurements were carried out in a ${}^{3}$He/${}^{4}$He Oxford dilution refrigerator at a base temperature of $\sim$20 mK. Four-terminal differential resistance and I-V measurements were performed simultaneously by superimposing a small ac voltage of frequency $\sim$10-100 Hz on a dc bias using a home-built summing amplifier and feeding the output to a home-built current source with a nominal output impedance of $\sim10^{13}\ \Omega$. The voltage across the junction was amplified at room temperature with a voltage preamp with $10^{9} \ \Omega$ input impedance, whose output was sent simultaneously to a dc voltmeter to measure the dc voltage and to a lock-in amplifier to measure the ac voltage. The summing amplifier, current source and first stage voltage amplifier were battery operated and were housed in a $\mu$-metal box.  Wiring to the sample at the mixing chamber consisted of twisted pairs of superconducting wires thermally anchored at each stage of the dilution refrigerator, with radio-frequency $\pi$ filters with a 3 dB cut-off frequency of 800 kHz placed at the top of the refrigerator to minimize heating from radio-frequency sources.  Special care was taken to minimize line frequency noise by eliminating ground loops. A low frequency spectrum was measured prior to each measurement to confirm that the noise level was dominated by the first stage voltage amplifier, with no or minimal line frequency contributions. AC excitations were small ($\sim25$ pA) to allow the full features of the differential resistance to be observed. 
\subsubsection{Atomic Force Microscopy}
Samples were analyzed using atomic force microscopy (AFM) (Asylum Research Cypher, Oxford Instruments) to determine feature dimensions in the coplanar waveguide resonator and the capacitor pads. The AFM images were processed using Gwyddion 2.62. Line profiles were drawn across the trenches to measure trench width and depth and across the capacitor pad gap to measure gap width and depth. Additionally, the surface roughness of both pads in each qubit was calculated.

For cryogenic AFM studies, a similar approach to that described previously was employed \cite{lee2021discovery}. An attoCube AFM system, incorporated with a Quantum Design 9T physical property measurement system (PPMS) was used to directly map the surface of superconducting qubit devices across a range of temperatures from room temperature to 2 K in a helium gas environment. AFM scanning in standard contact mode using a nanosensors pointprobe plus tip allows imaging of a 38×38 $\mu$m field of view.

\subsection*{Qubit Measurement Setup and Method} \label{measurements}
Details regarding the microwave package, microwave shielding environment and various measurement setup parameters is provided in Ref. \cite{RN1}. The qubit state was measured via dispersive readout and T$_1$ measurements were performed continuously for 10 hours for each qubit.

\begin{figure*}[t]
  \begin{center}
    \includegraphics[width= 7in]{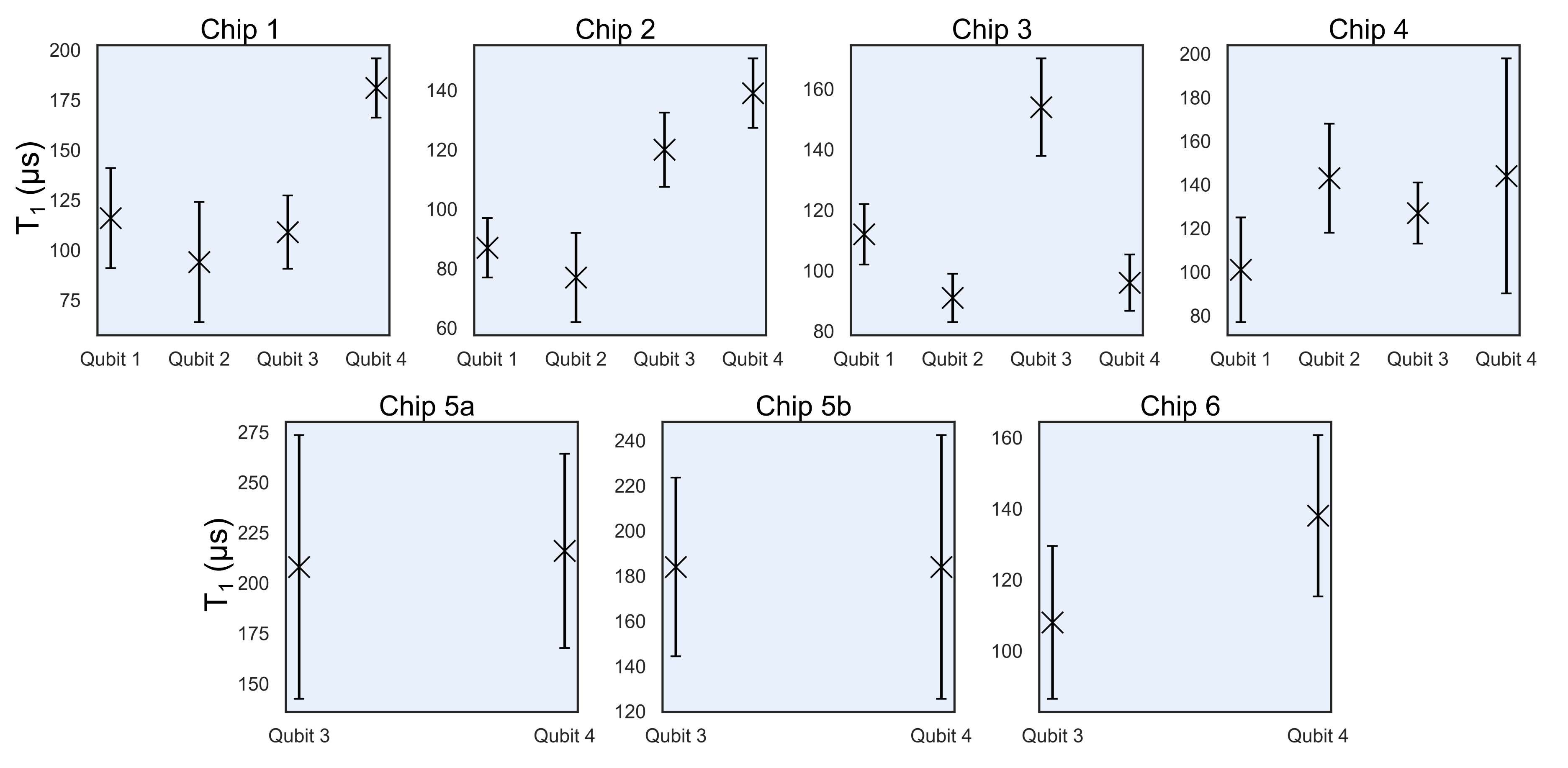}

    \caption{Spread in T$_1$ measurements for 22 qubit devices measured across 8 chips. Measurements were performed continuously for 10 hours for each qubit. Error bars represent $\pm 1\sigma$.}
  	\label{fig:SQubit}
  \end{center}
\end{figure*}

22 qubit devices were measured across 7 chips. The resultant average and maximum T$_1$ values as well as average quality factors, $Q$, are reported in Table 1. $Q$ is calculated per the following relation, $Q=2\pi*f*T_1^{avg}$, where $f$ represents qubit frequency and T$_1^{avg}$ represents the average T$_1$ for a given qubit. The spread in T$_1$ measurements is provided in Figure \ref{fig:SQubit}.

\clearpage
\newpage

\begin{figure*}[t]
  \begin{center}
    \includegraphics[width= 7in]{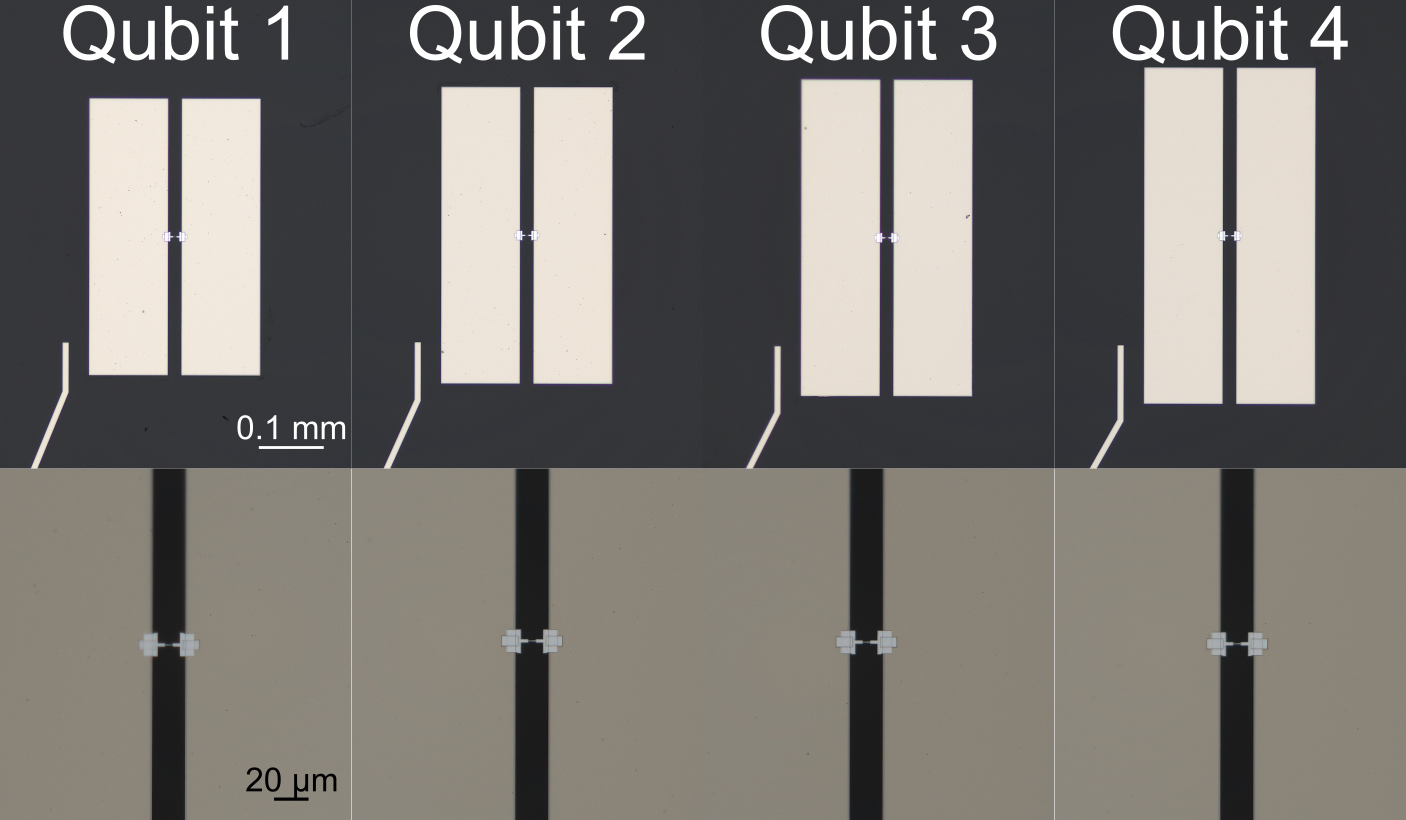}

    \caption{Representative optical images of qubits 1-4. The length of the rectangular pads are tuned across the 4 devices to minimize frequency collisions. The scale bar is consistent across each row.}
  	\label{fig:S1}
  \end{center}
\end{figure*} 

\begin{figure*}[t]
  \begin{center}
    \includegraphics[width= 7in]{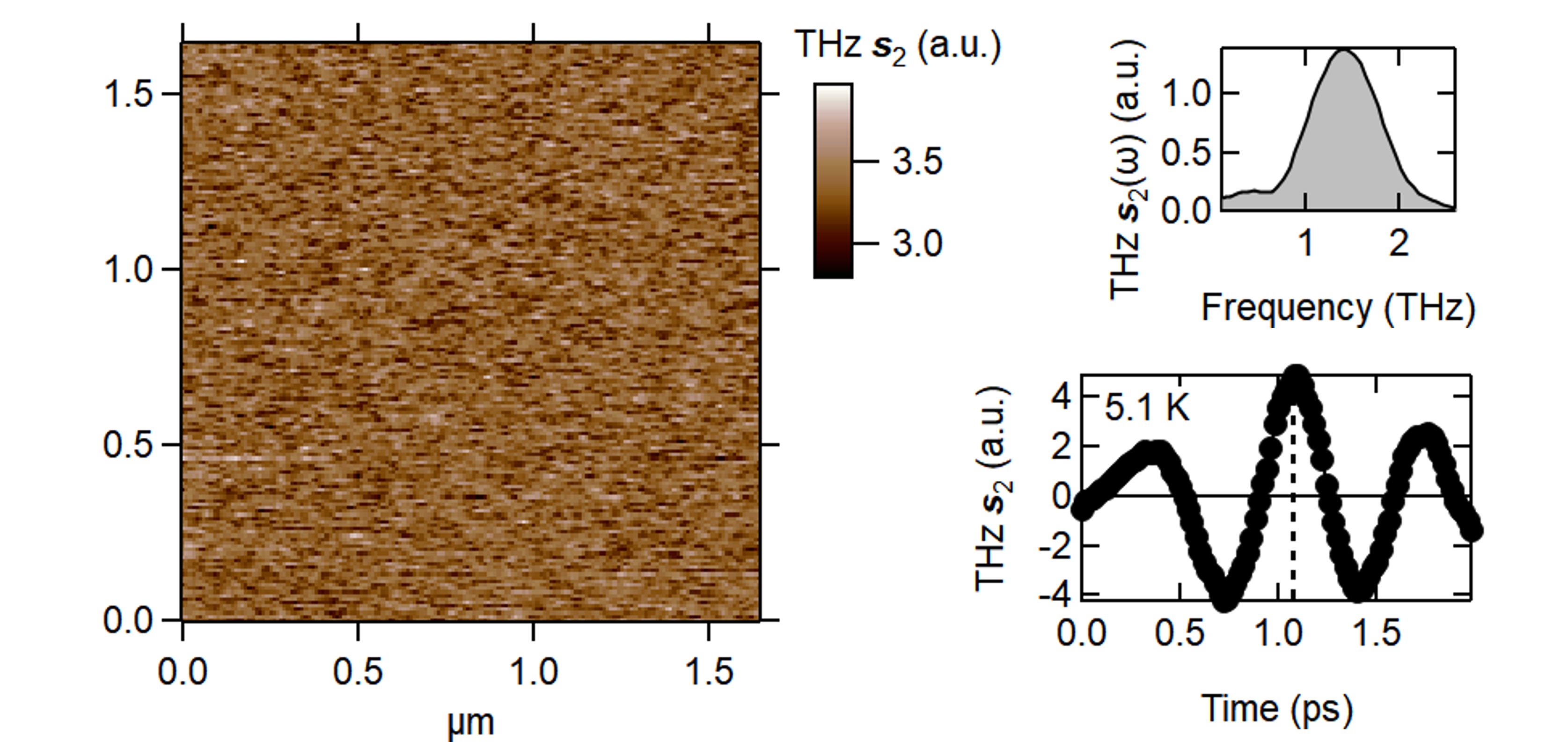}

    \caption{A 1.65 $\mu$m x 1.65 $\mu$m area of a THz near-field image of DC high power sputtered Nb film of 175 nm nominal thickness deposited on high-resistivity Si wafer at 5.1 K temperature below the T$_c$ of Nb. A time trace taken at a single point and its fast Fourier transformed spectrum for THz nanospectroscopy is shown on the right.}
  	\label{fig:STHz}
  \end{center}
\end{figure*} 

\begin{figure*}[t]
  \begin{center}
    \includegraphics[width= 7in]{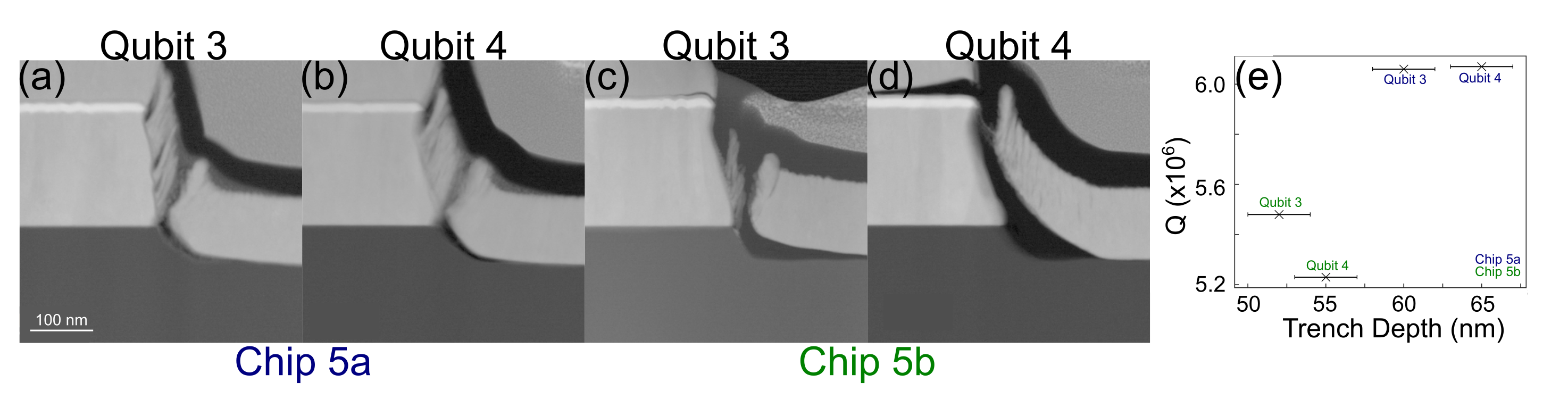}

    \caption{(a-d) Electron microscopy images taken of the sidewall geometry for qubits 3 and 4 on Chip 5a and for qubits 3 and 4 on Chip 5b consisting of a base metal of Nb film capped with Ta. (e) Trend between trench depth and qubit quality factor for the aforementioned 4 qubits. The scale bar is consistent for (a-d)}
  	\label{fig:S2}
  \end{center}
\end{figure*} 

\begin{figure*}[t]
  \begin{center}
    \includegraphics[width= 3.5in]{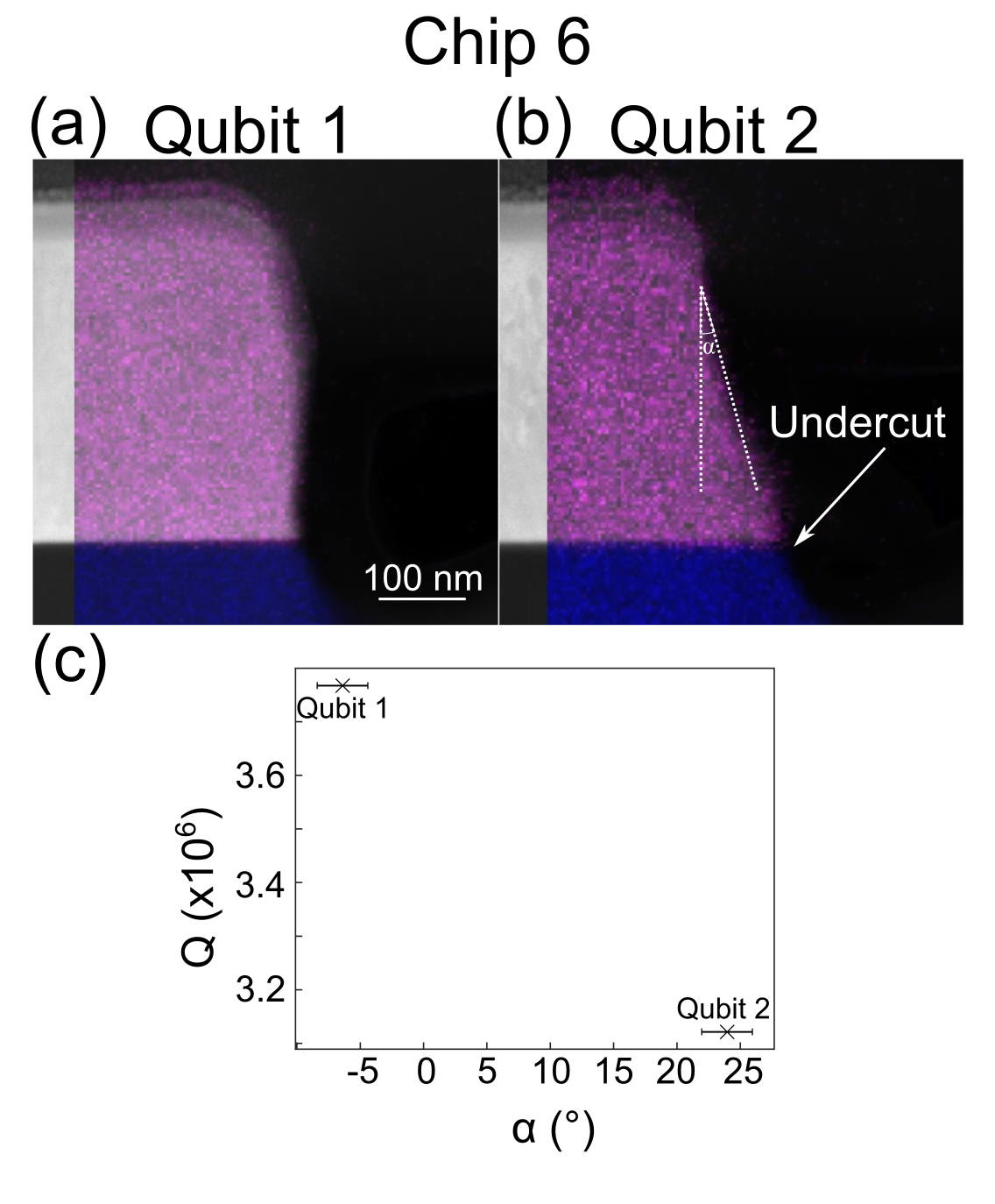}

    \caption{Electron microscopy images taken of the sidewall geometry for qubits 3 and 4 on Chip 6 consisting of a base metal of Nb film. Nb and O EDS maps are provided as an overlay (e) Trend between $\alpha$ as defined in Figure 4 and qubit quality factor for these two qubits. The scale bar is consistent for (a-b).}
  	\label{fig:S3}
  \end{center}
\end{figure*} 

\begin{figure*}[t]
  \begin{center}
    \includegraphics[width= 7in]{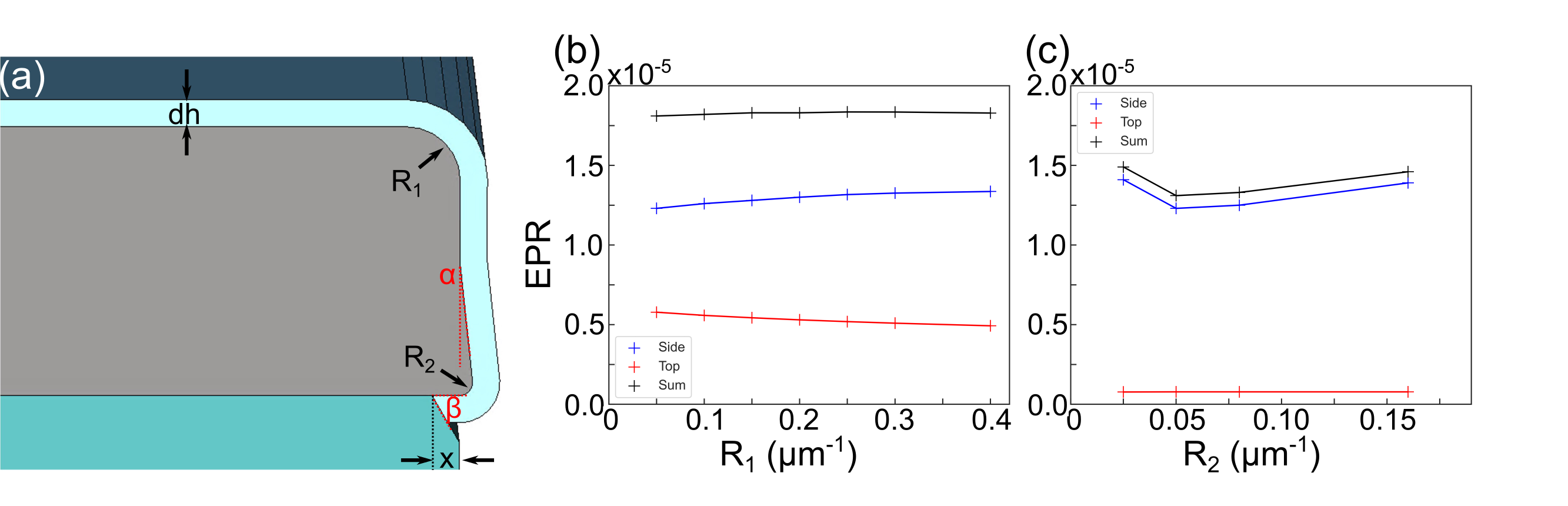}

    \caption{EM simulations of the qubit sidewall. Sidewall surface, top surface, and total surface EPR values are provided as a function of (b) R$_1$ and (c) R$_2$. We observe minimal changes in the total surface EPR as R$_1$ and R$_2$ are varied.}
  	\label{fig:S4}
  \end{center}
\end{figure*}

\begin{figure*}[t]
  \begin{center}
    \includegraphics[width= 7in]{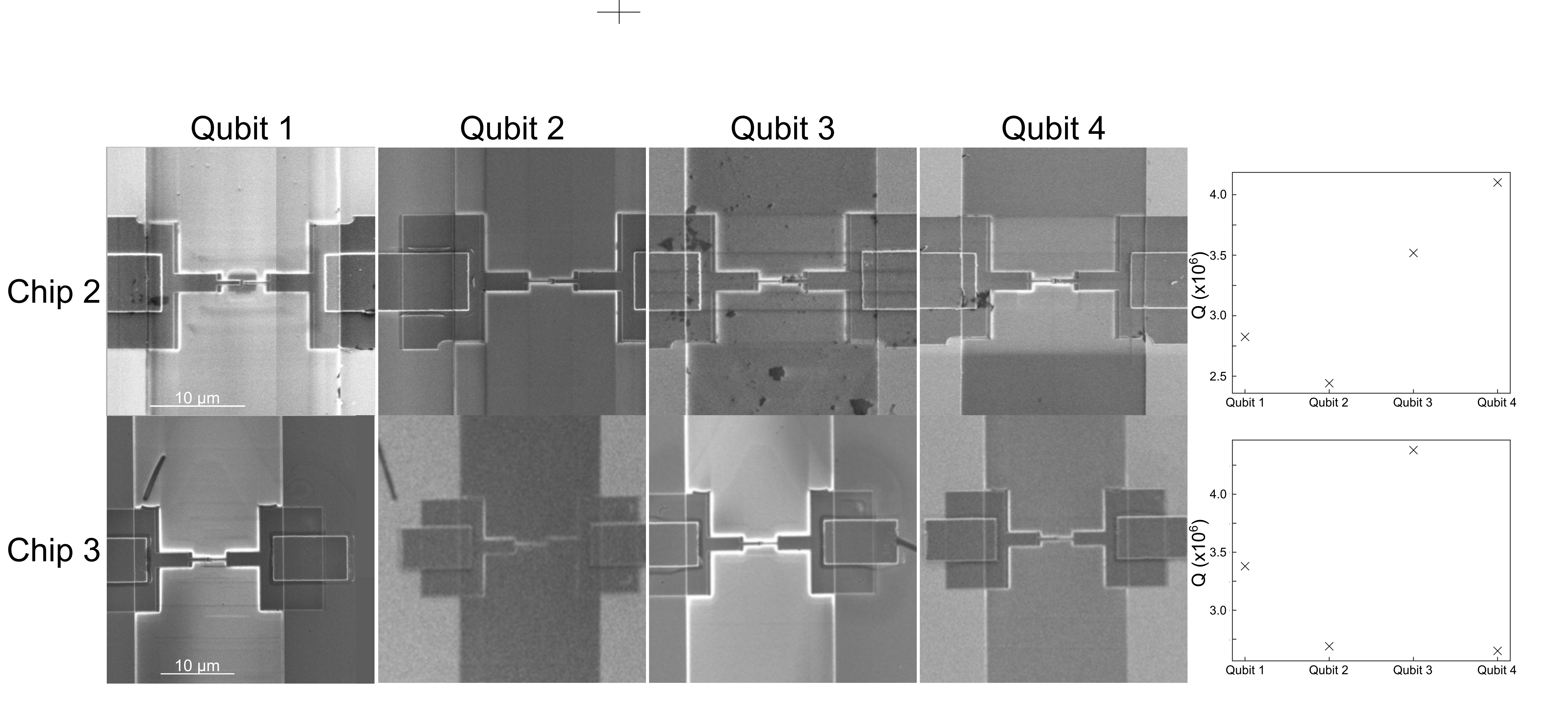}

    \caption{Scanning electron microscopy images taken of the junction area for qubits 1-4 on Chips 2 and 3. Macroscopic debris is observed in many of the qubits, but we observe no trend between particle density and qubit quality factor. The scale bar is consistent across each row.}  	\label{fig:S5}
  \end{center}
\end{figure*} 

\begin{figure*}[t]
  \begin{center}
    \includegraphics[width= 7in]{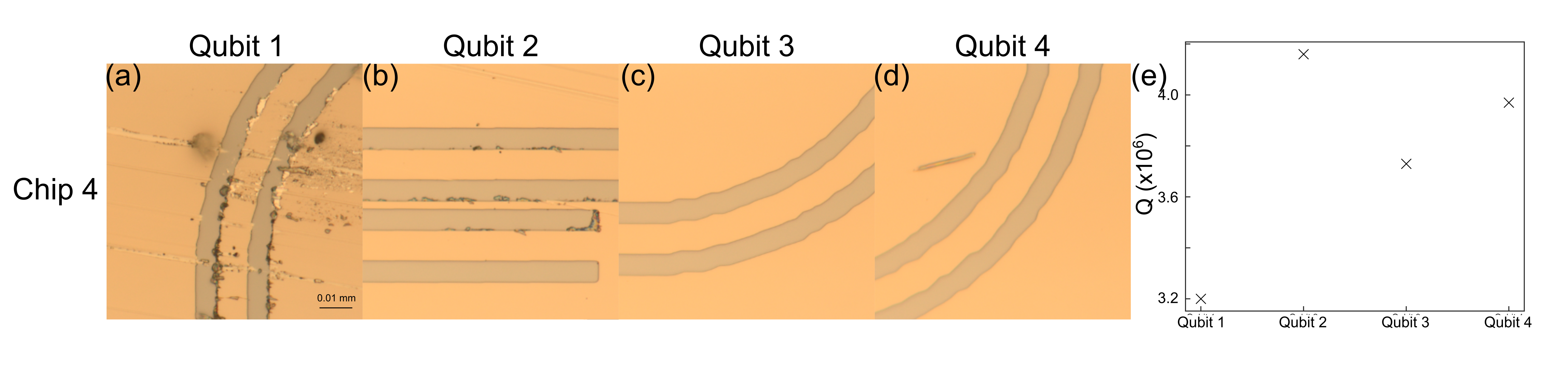}

    \caption{Optical microscopy images taken from the readout resonator region for qubits 1-4 on Chip 4. Lithographic defects are observed in many of the resonator regions, but we observe no trend between such defects and qubit quality factor. The scale bar is consistent for (a-d)}
  	\label{fig:S6}
  \end{center}
\end{figure*} 

\begin{figure*}[t]
  \begin{center}
    \includegraphics[width= 7in]{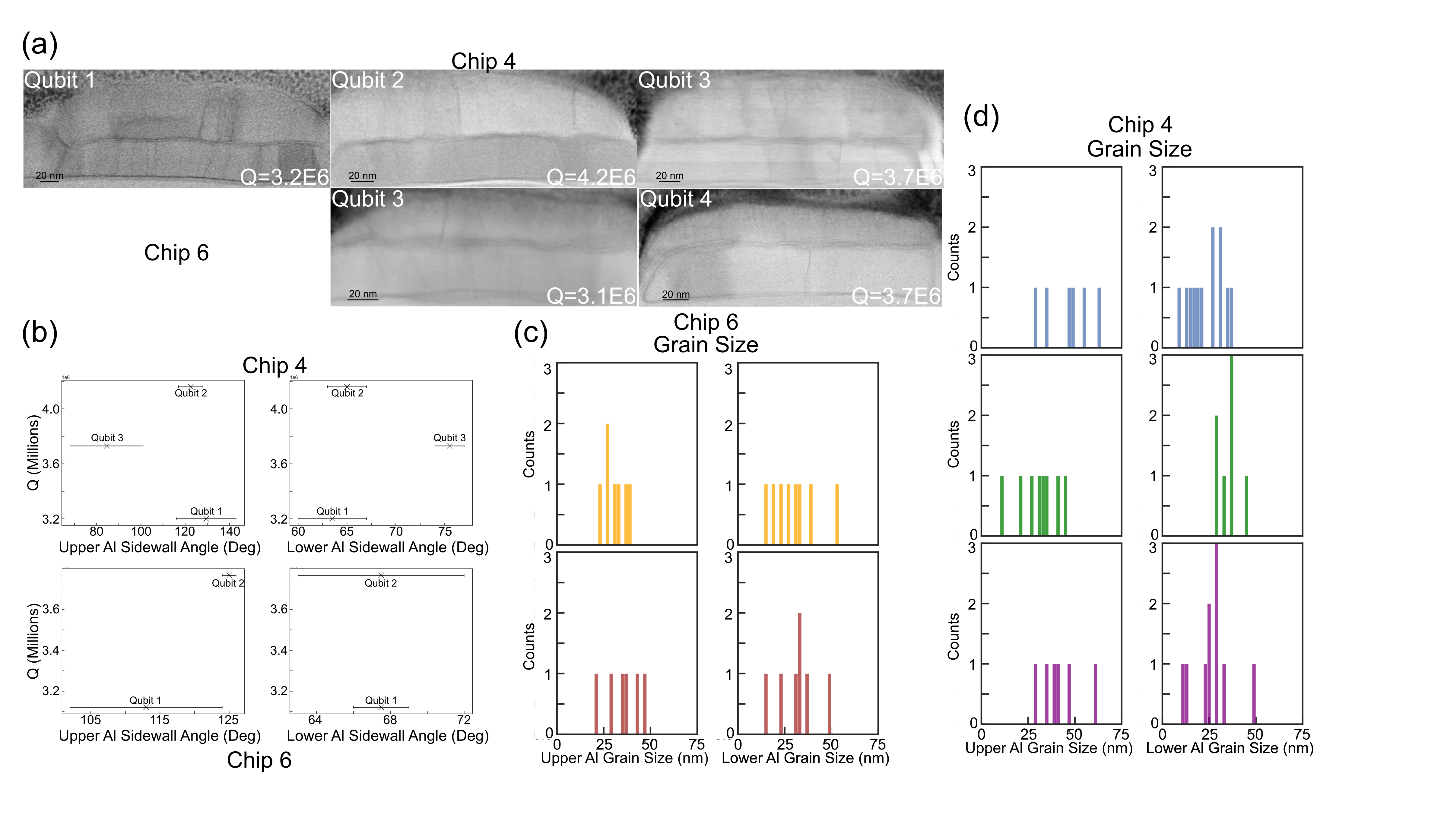}

    \caption{Structural characterization of the Josephson junction region in Chips 4 and 6. Concerning the Josephson Junction region, we do not find any correlation between sidewall angle or grain size on qubit performance.}  	\label{fig:S7}
  \end{center}
\end{figure*} 

\begin{figure*}[t]
  \begin{center}
    \includegraphics[width= 7in]{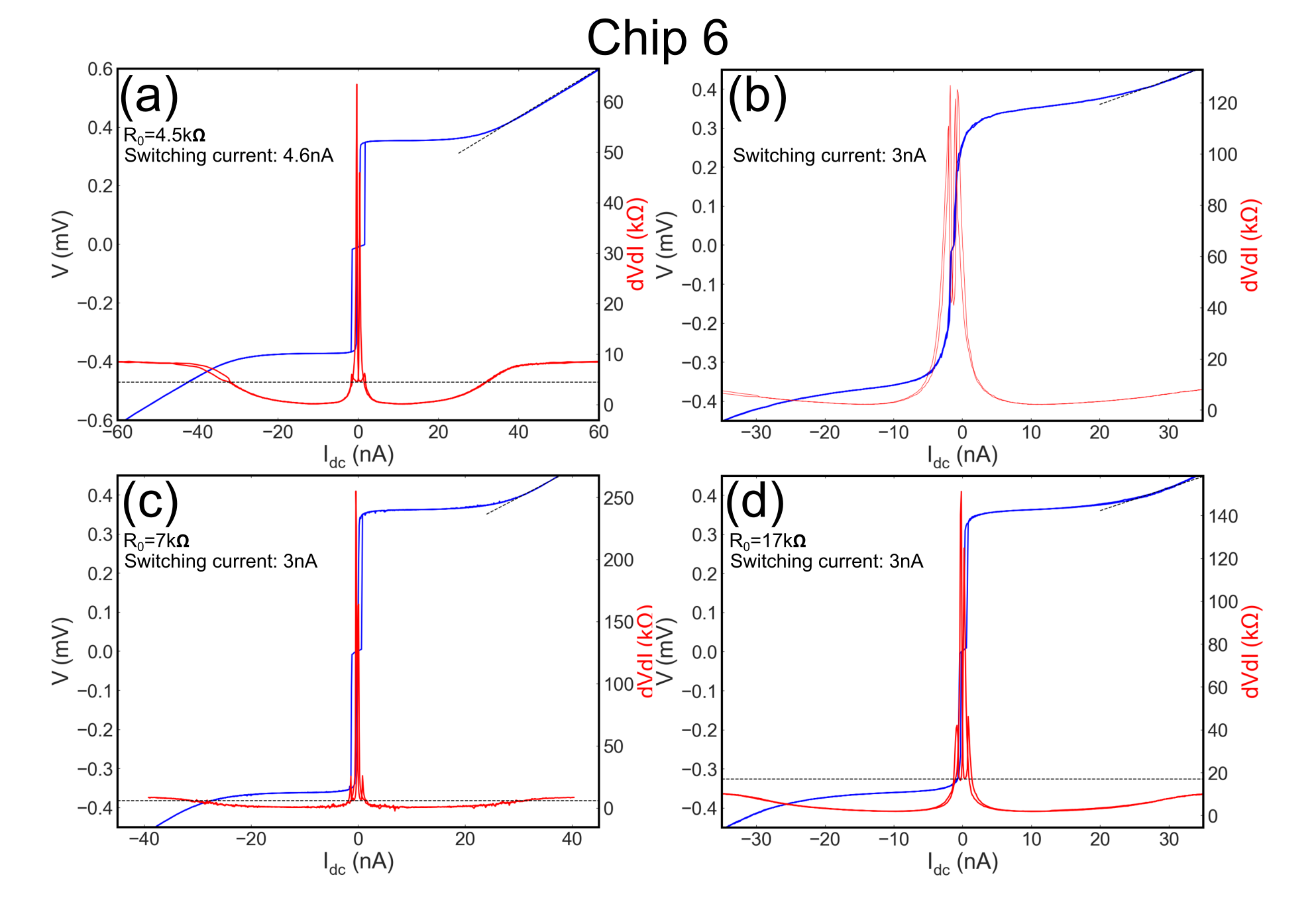}

    \caption{I-V measurements taken across the Josephson Junction region in Chip 4. Although we observe variations in normal state resistance and switching current from qubit-to-qubit, we do not observe any trend with performance.}
  	\label{fig:S8}
  \end{center}
\end{figure*} 

\begin{figure*}[t]
  \begin{center}
    \includegraphics[width= 7in]{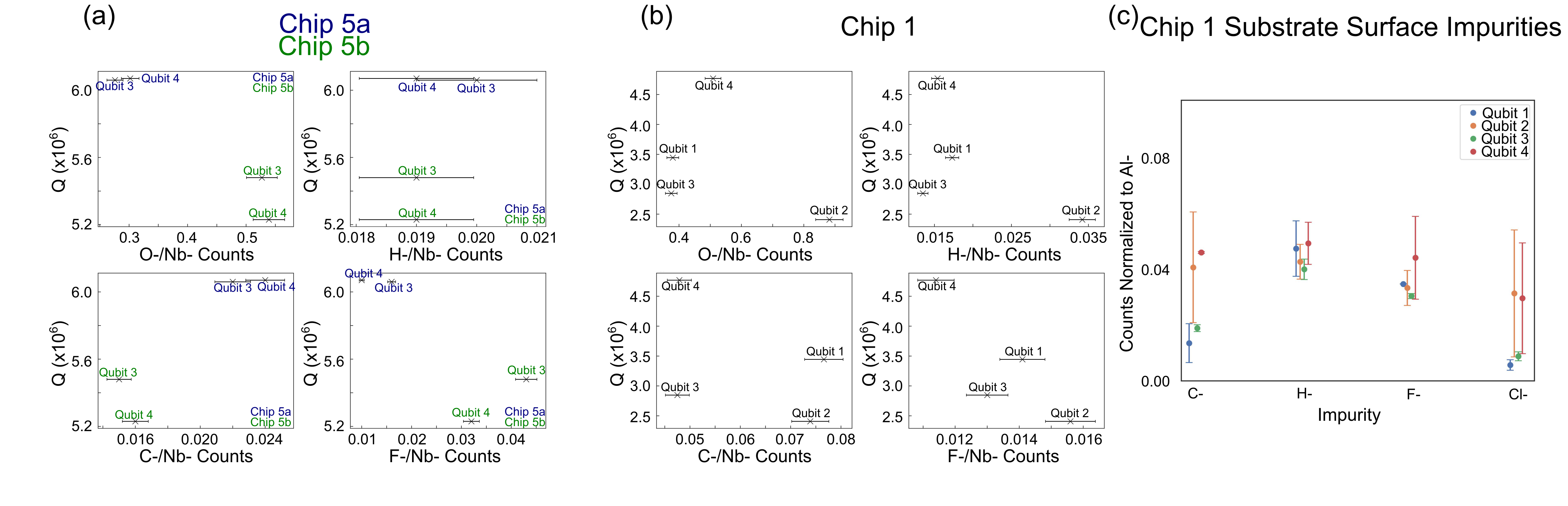}

    \caption{ToF-SIMS was employed to probe the surface chemistry of the superconducting pads as well as the exposed substrate surface. The average impurity counts for qubits across 3 chips is provided here. There do not appear to be any obvious trends between the surface chemistry of the superconducting pads or the substrate and the qubit performance.}
  	\label{fig:S9}
  \end{center}
\end{figure*} 

\begin{figure*}[t]
  \begin{center}
    \includegraphics[width= 7in]{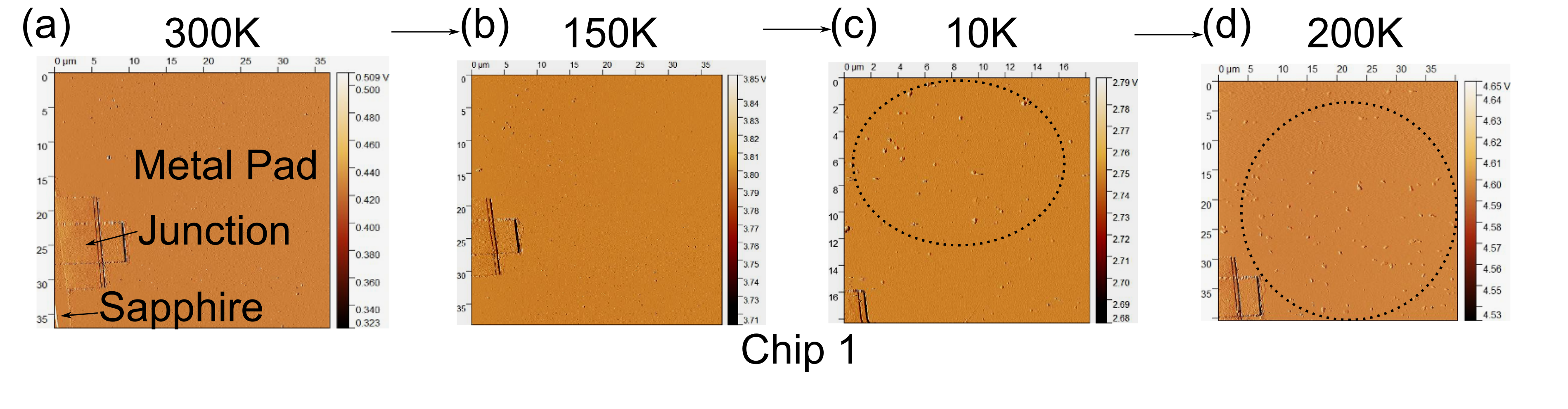}

    \caption{AFM surface images of the superconducting pad for Chip 1 are shown as a function of temperature. The formation of niobium hydride precipitates is observed at T=10 K and remains stable up to T=200K.}
  	\label{fig:S10}
  \end{center}
\end{figure*} 

\begin{figure*}[t]
  \begin{center}
    \includegraphics[width= 7in]{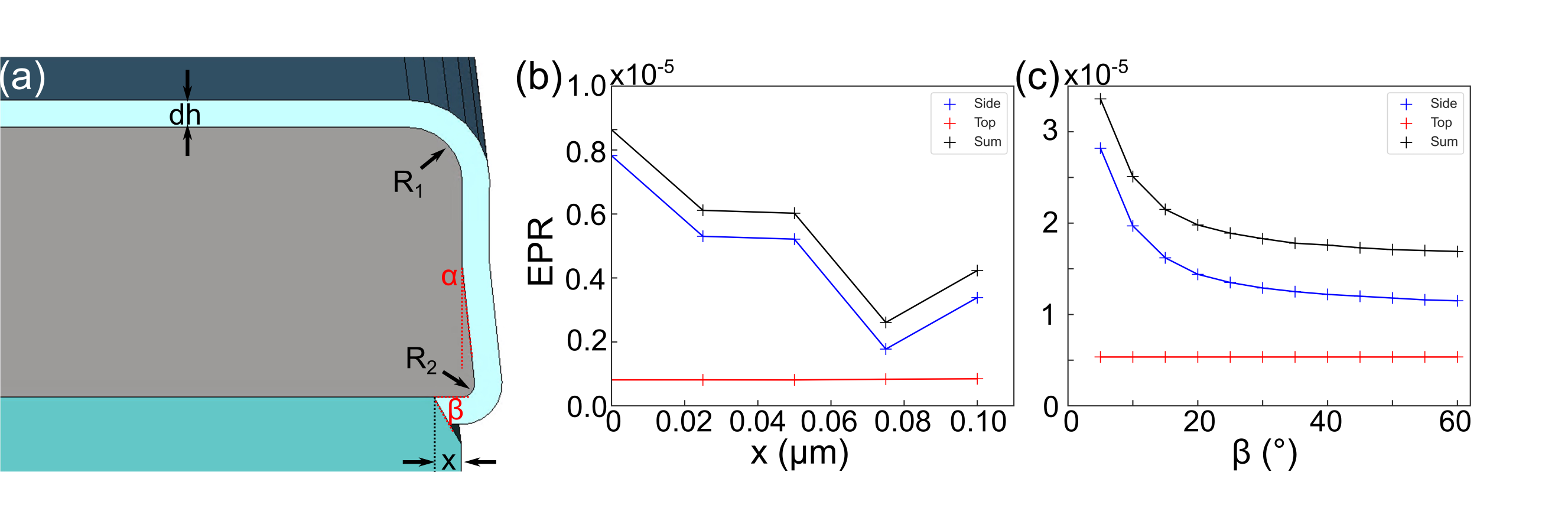}

    \caption{EM simulations of the qubit sidewall with a focus on exploring the effect of a substrate undercut. Surface EPR values are provided as a function of (b) x and (c) $\beta$. We observe that an undercut on the order of 60 nm can effectively reduce the surface EPR by 2$\times$. At the same time, careful consideration must be given to the undercut angle, as shallow angles can trap EM fields and greatly increase the total EPR.}
    \label{fig:S11}
  \end{center}
\end{figure*} 

\end{document}